\documentclass[aps,prb,amssymb,twocolumn,showpacs]{revtex4}
\usepackage{graphicx}
\usepackage{times}

\newcommand{\bra}[1]{\langle #1 |}
\newcommand{\ket}[1]{| #1 \rangle}

\newcommand{\average}[1]{ \langle #1 \rangle}

\newcommand{\be}{\begin{equation}}
\newcommand{\ee}{\end{equation}}
\newcommand{\bae}{\begin{eqnarray}}
\newcommand{\eae}{\end{eqnarray}}

\begin{document}

\title{Quantum discord and classical correlations in the bond-charge Hubbard model: quantum phase transitions, ODLRO and violation of the monogamy property for discord}

\author{Michele Allegra$^{1}$, Paolo Giorda$^{1}$ and Arianna Montorsi$^{2}$}

\address{$^1$Institute for Scientific Interchange (ISI), Villa Gualino, Viale Settimio Severo 65, I-10133 Torino, Italy}

\address{$^2$Dipartimento di Fisica del Politecnico and CNISM, Corso Duca degli Abruzzi 24, I-10129 Torino, Italy}

\pacs{03.65.Ud, 03.67.-a, 71.10.Fd}

\begin{abstract}
We study the quantum discord (QD) and the classical correlations (CC)  in a reference model for strongly correlated electrons, the one-dimensional bond-charge extended Hubbard model. We show that the comparison of QD  and CC and of their derivatives in the direct and reciprocal lattice allows one to efficiently inspect the structure of two-points driven quantum phase transitions (QPTs), discriminating those at which  off diagonal long range order (ODLRO) is involved. Moreover we observe that QD between pair of sites is a monotonic function of  ODLRO, thus establishing a direct relation between the latter and two point quantum correlations different from the entanglement. 
The study of the ground state properties allows to show that for a whole class of permutation invariant ($\eta$-pair) states quantum discord can violate the \textit{monogamy} property, both in presence and in absence of bipartite entanglement. In the thermodynamic limit, due to the presence of ODLRO, the violation for $\eta$-pair states is maximal, while for the purely fermionic ground state is finite. From a general perspective, all our results validate the importance of the concepts of QD and CC for the study of critical condensed-matter systems.

\end{abstract}
\maketitle
\section{Introduction}
\label{Sec.: introduction}

A very fertile interplay between the theories of quantum-information~\cite{QuantumInfo} and condensed matter~\cite{CondMat} has developed during the last decade.
On one side, condensed-matter theory has suggested a wide range of possibilities for the implementation of quantum communicational~\cite{Bose} and computational~\cite{Kane} tasks. On the other side, quantum information theory has provided novel and deep insights into the physics of condensed-matter systems.
In particular, since the capability of entanglement to mark quantum phase transitions (QPT) was first recognized~\cite{Osterloh, Osborne}, the concept of quantum correlations has become essential in characterizing quantum critical phenomena~\cite{Vedralrev,Amico,Latorre, Rot, Wu1, Gu}. In this context, \textit{quantum correlations} and \textit{entanglement} have been usually identified as one and the same concept. It is now accepted that the notion of entanglement~\cite{HOR4} is unfit to capture the whole amount of quantum correlations present in a system. The introduction of quantum discord (QD)~\cite{Ollivier, Vedral}  in fact showed that a system can exhibit quantum correlations even in absence of entanglement. Furthermore, QD allows to properly distinguish the total correlations between two subsystems in terms of quantum and classical~\cite{Vedral}. QD was first devised in the realm of quantum-measurement theory~\cite{Ollivier} but it has been analyzed in a large variety of different physical systems. In a quantum information perspective, it has been suggested that
QD, rather than entanglement, may be the most fundamental resource allowing for the speedups of quantum over classical computation~\cite{Datta2}, and that it may have a relevant role in quantum communications protocols~\cite{QComm}. The study of randomly generated states has established that QD is present in almost all quantum states~\cite{Ferraro}. The notion of QD has also been investigated, both from the theoretical~\cite{opendynamics-M,opendynamics-nM} and the experimental side~\cite{opendynamics-exp}, in the context of open quantum systems where it was shown that QD is generally more robust than entanglement to dissipation and thermal noise; moreover, QD is tightly related to complete positivity of quantum maps~\cite{Shabani}.
In the realm of many-body systems, the behavior of QD has been analyzed in relation to QPTs and thermal effects~\cite{Dillen, Sarandy, Werlang, Maziero,Yurischev, Chen1,Maziero2}. So far, the research has mostly concentrated on one-dimensional spin $1/2$ models ~\cite{Dillen, Sarandy, Werlang, Maziero, Yurischev,Maziero2}.
The main results of these analyses show that two-point QD and CC between near as well as distant sites show clear signatures of QPTs (discontinuities or divergences), which can be understood within a general framework~\cite{Dillen} and agree with finite-size scaling theory in the case of finite chains.
Apart from spin systems, many-body QD has been studied in the LMG model~\cite{Sarandy} and the Castelnovo-Chamon model~\cite{Chen1}, where it marks a topological QPT.\\
At present, a thorough analysis of QD and CC in correlated electron systems is still lacking, mainly because the latter, at variance with the simplest spin systems, requires the evaluation of the discord for pairs of q-dits. The present work is intended to fill this gap, by investigating the behavior of QD and CC for the ground states of the one-dimensional bond-charge extended Hubbard model~\cite{Sim,Arrachea}, which is a reference model in correlated-electron theory. The model  has an integrable point, and its entanglement properties  have been the subject of recent studies ~\cite{Anfossi1, Anfossi2, Anfossi3,Giordak,Vitoriano} where use of  two-point and multipartite entanglement measures led to a classification of QPTs into multipartite or two-point driven.  These studies left open the problem of addressing the general role of bipartite correlations for all two-points driven QPTs, as well as their relation with the presence of off diagonal long range order (ODLRO) which characterizes some ordered phases of  the model. The introduction of QD and CC allows to solve this problem in a proper framework.\\
We systematically consider the quantum discord and the classical correlations, in direct space between two-sites and in momentum space between two couples of modes, and we study their interplay and their ability to properly describe the rich zero temperature phase diagram and the various phase transitions exhibited by the extended Hubbard model we consider. We see that QD and CC can highlight the presence of a so-called entanglement transition, where a different role of quantum and classical correlation at a transtion is revealed by the different behaviour of QD and CC, both in their maxima and in the divergence of their derivatives (subsection \ref{Sec.: regionI}). The study of the derivatives of QD and CC close to the critical lines allows to confirm the two-point/multipartite nature of the various transitions and to distinguish transitions that are physically different based on a different role of long range quantum correlations(ODLRO) (subsection \ref{Sec.: regionII}). We demonstrate that these long-range correlations correlations, which are at the basis of superconductivity, are related to two-point discord rather than two-point entanglement: indeed a direct relation between ODLRO and QD can be found (subsection \ref{Sec.: regionIII}). 
This relation is true both in the direct and in the reciprocal lattice picture, since a functional relation between the two-site QD and the two mode QD can be established (subsection \ref{Sec.: reclattice}). As an example of how condensed matter systems constitute a natural playground to test quantum information concepts, our study of the ground state properties also  sheds light onto an aspect of the quantum correlations that is very relevant in the general context of quantum information theory: the monogamy property (we address this issue in detail in the self-contained subsection~\ref{Sec.: monogamy}). Upon considering ground states of the model also at finite system size, we can extend previous analyses of the monogamy relation to an $n$-partite setting with $n \geq 3$. In a phase of the model the ground states coincide with a class of permutation-invariant states, for which we show that the monogamy relation is always violated, both in presence and in absence of entanglement. In the TDL the entanglement vanishes and the violation of the monogamy property for QD becomes maximal: due to the presence of ODLRO, a single qubit can exhibit finite amount of discord with an infinite number of other qubits. The monogamy relation can be violated also in absence of ODLRO, but in this case the violation is not maximal.\par
The paper is organised as follows.
Section~\ref{Sec.: discord} is an introduction to QD. Section~\ref{Sec.: Hubbard model} is a brief review of the main features of the extended Hubbard model: Hamiltonian, phase diagram., etc.
Section~\ref{Sec.: results} is the core of the work, displaying our analysis of the behavior of QD and CC in the whole phase diagram of our model with a special focus on quantum critical points/lines.  
Section~\ref{Sec.: conclusions} closes the paper, highlighting and summarizing the main conclusions of our work.

\section{Quantum discord and classical correlations}
\label{Sec.: discord}

The peculiarity of quantum physics mirrors into the highly nonclassical nature of correlations between quantum systems. It is widely known that quantum entanglement~\cite{HOR4} lays at the heart of many quantum phenomena and plays a crucial role in quantum information processing. However, entanglement alone is insufficient to describe the quantum character of the correlations present in quantum states. The definition of a proper measure of quantum correlations can be derived as the difference between the total correlations between two subsystems $A$ and $B$, represented by the quantum mutual information $I(A:B)$, and the classical correlations $C(A:B)$, whose measure has been introduced in \cite{Vedral}:
\be
Q(A:B)=I(A:B) - C(A:B) \label{Eq.: DefDiscord}
\ee
where $I(A:B) = S(A) + S(B) - S(A:B)$ is expressed in terms of the von Neumann entropy of reduced density matrices $\rho_A, \rho_B$ and the density matrix of the system $\rho$ respectively. The measure for classical correlations is defined as the maximal amount of information on one of the subsystems, say $A$, that one can extract by classical means i.e.,  by operating a complete measurement process the other subsystem $B$
\be
C(A:B) = \max_{B_k} [S(A) - \sum_k p_k S(\rho_{Ak})].\label{Eq.: DefClassicalCorrs}
\ee
Here the set of positive operators $\{\openone_A\otimes B_k\},\ \ \sum_k B_k = \openone_B$ represent a von Neumann measurement, i.e., a set of orthogonal projectors on subsystem $B$ and $\rho_{k} = \mbox{Tr}_{B}[ \frac{1}{p_k} B_k \rho B_k$] is the $k$-th post-measurement state of subsystem $A$ obtained with probability $p_k = \mbox{Tr}_{A}\rho_{k} $ (see the Appendix for details). The maximum over all von Neumann measurements is attained in correspondence of the minimum of the conditional entropy $S(A|B) = \sum_k p_k S(\rho_{Ak})$ and in general requires difficult optimization procedures. The definition of $Q(A:B)$ above coincides with the definition of quantum discord originally given in~\cite{Ollivier}. The reasons that led to this definition and the properties of this measure explain how it entails for the quantification of quantum correlations. The condition that captures the nature of the quantum correlations described by the discord is that if a state has non zero discord all complete measurements on a subsystem $B$ will unavoidably disturb the whole system and in particular the subsystem $A$:
\be
\rho\neq \sum_k \openone_A\otimes B_k \rho \openone_A\otimes B_k
\ee
This is a quantum feature that has no classical counterpart, but it does not require the presence of entanglement: separable states can have finite amount of discord.
The discord $Q(A:B)=0$ iff the state $\rho$ is diagonal in a product eigenbasis~\cite{Datta1}, and thus a classical state has the form $\{\ket{i}_A\otimes\ket{j}_B\}$ i.e., $\rho=\sum_{i,j} \lambda_{i,j}\ket{i}_{AA}\bra{i}\otimes\ket{j}_{BB}\bra{j}$, for a possible classification of quantum states see \cite{AliQuantumStates}. Quantum discord is a correlation measure aimed at quantifying all quantum correlations including entanglement. A key feature of QD is that in general it is not symmetric under exchange of the two subsystems~\cite{Maziero3}: measurement processes applied to the two different subsystems can lead to different values of discord. However, in the examples we will study the density matrices are symmetric with respect to the exchange of subsystem and so is the discord evaluated for the systems represented by these states. As already mentioned, the evaluation of QD and CC poses in general some difficulties since it requires an optimization procedure.
For two qubit systems analytical formulas have been found only for special examples of two qubit density matrices \cite{Luo,Ali,QChen,Girolami}, while for continuous variable systems a general formula have been derived for Gaussian states only \cite{GiordaGaussDiscord}. In the following we will use the analytical formulas~\cite{Ali} for the two qubit case, while we will resort to a simple numerical optimization for the case of two qutrits.

\section{The bond-charge extended Hubbard model}
\label{Sec.: Hubbard model}

\subsection{Basics of the model}

The bond-charge extended Hubbard model was derived as an effective one-band Hamiltonian for the description of cuprate superconductors~\cite{Sim}. The model is described by the following Hamiltonian:
\begin{eqnarray}
    H_{BC} =  &-& \sum_{<i,j>\sigma}[1 - x (n_{i \bar{\sigma}}+n_{j \bar{\sigma}})]
    c_{i \sigma}^\dagger c_{j \sigma} -\mu\sum_{i\sigma}n_{i\sigma}\nonumber\\
    & & + u \sum_i \left( n_{i \uparrow}-\frac{1}{2}\right) \left(n_{i \downarrow}-\frac{1}{2}\right)  \label{hubbardham}
\end{eqnarray}
where $c_{{i} \sigma}^\dagger$ and $c_{{i} \sigma}^{} \,$ are fermionic creation and annihilation operators on a one-dimensional chain of length $L$; $\sigma = \uparrow, \downarrow$ is the spin label, $\bar{\sigma}$ denotes its opposite, ${n}^{}_{j \sigma} = c_{j \sigma}^\dagger c_{j \sigma}^{}$ is the spin-$\sigma$ electron charge, and $\langle {i} , \, {j} \rangle$ stands for neighboring sites on the chain; $u$ and $x$ ($0\leq x \leq 1$) are the (dimensionless) on-site Coulomb repulsion and bond-charge interaction parameters; $\mu$ is the chemical potential, and the corresponding term allows for arbitrary filling.
\begin{figure}[!h]
   \begin{center}
        \fbox{\includegraphics[height=5cm, width=4.2 cm, viewport= 10 20 280 230,clip]{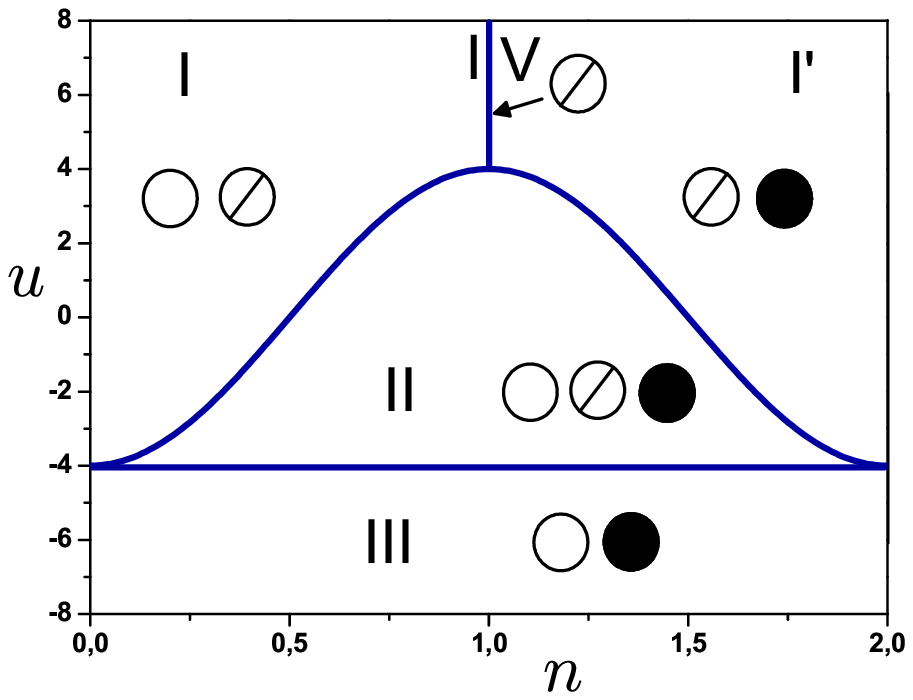}
        \includegraphics[height=5cm, width=4.2 cm,viewport= 10 12 280 217,clip]{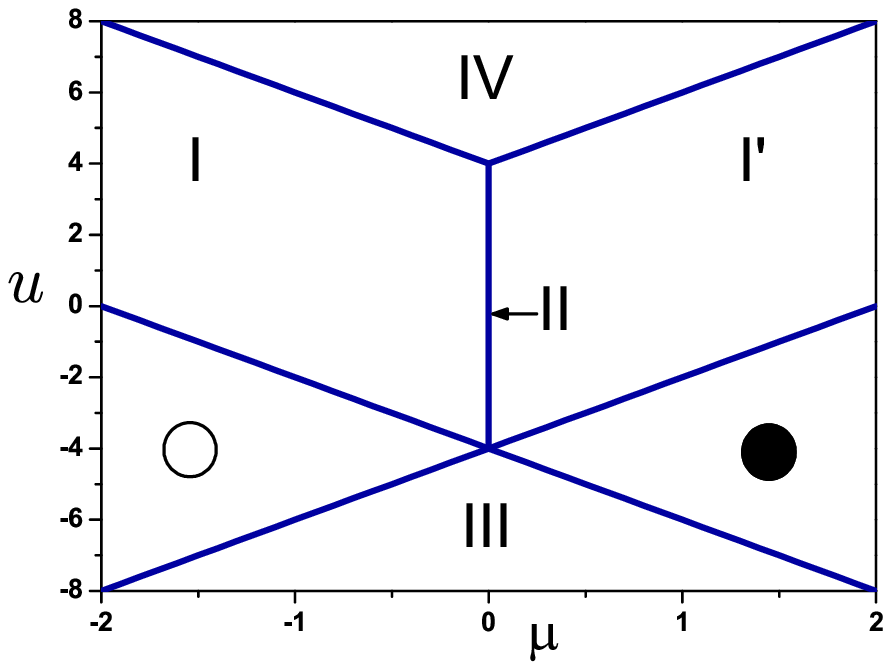}}
        \caption{Ground-state phase diagram of $H$. LEFT: $n$-$u$
         plane; empty circles stand for empty sites, slashed and full
         circles stand for singly and doubly occupied sites, respectively. RIGHT:
         $\mu$-$u$ plane.}\label{FGS1}
    \end{center}
\end{figure}

The model is considered here at $x=1$, in which case the system becomes integrable. This happens for two reasons. First, the $1 - x (n_{i \bar{\sigma}}+n_{j \bar{\sigma}})$ term suppresses several hopping possibilities.
As a result, we can separate the four possible states at site $i$ into two groups, namely $A = \{\ket{\uparrow}, \ket{\downarrow}\}$ and $ B = \{\ket{0}, \ket{\uparrow \downarrow}\}$: hopping permutes states of group $A$ with states of group $B$, but not states of the same group. The role of spin orientation becomes dynamically irrelevant, and the system behaves as if at each site the local space had dimension $3$: $\ket{\uparrow}$ and $\ket{\downarrow}$ can be considered as the same state. Second, the hopping term commutes with the terms in $u, \mu$ and the number of doubly occupied sites becomes therefore a conserved quantity. \\
The physics of the system described by $H$ is basically that of $N_s$ spinless fermions - singly occupied sites - which move in a background of $L-N_s$ bosons, of which $N_d$ are doubly occupied sites and the remaining are empty sites . Both $N_s$ and $N_d$ are conserved quantities, and determine the total number of electrons $N=N_s+2 N_d$. \\
The situation may be understood in the formalism developed by Sutherland in Ref. ~\cite{Sutherland}. We can say that, apart from constant terms, $H$ acts as a permutator of just two {\it Sutherland species} (SSs), the $N_s$ fermions, and the $L-N_s$ bosons. In practice, empty and doubly occupied sites ---though different as physical species--- belong to the same SS, since the off-diagonal part of the Hamiltonian (i.e., the hopping term) cannot distinguish between them. It is only the constant term counting doubly occupied sites which depends on the actual value of $N_d$. \\
It is convenient to rewrite both the Hamiltonian and the local vector space in terms of the Hubbard-like projection operators $X^{\alpha\beta}_i\doteq\ket{\alpha}_i\bra{\beta}_i$, with local algebra $X^{\alpha\beta}_i X^{\gamma\delta}_i = \delta_{\beta\gamma}X^{\alpha\delta}$ and nonlocal (anti-)commutation relations given by
\begin{equation}\label{HubProj}
    X^{\alpha\beta}_i X^{\gamma\delta}_j=(-)^{(\alpha+\beta) (\gamma+\delta)}
    X^{\gamma\delta}_j X^{\alpha\beta}_i \; , \; i\neq j \, ;
\end{equation}
here $\alpha=0,1,2$, $\ket{0}_i\equiv\ket{\mbox{vac}}_i$ is the local vacuum, $\ket{1}_i\doteq X^{10}_i \ket{0}_i$ is the singly occupied state (with odd parity), and $\ket{2}_i\doteq X^{20}_i \ket{0}_i$ is the doubly occupied state.
More precisely, as far as the ground state is concerned, the model Hamiltonian in the one-dimensional case can be fruitfully written as
\begin{eqnarray}
    H= &-& \sum_i \left (X^{10}_i X^{01}_{i+1}-X^{21}_i X^{12}_{i+1}
    + \mbox{H.c.}\right ) +u\sum_i X^{22}_i\nonumber \\
    &-&\left(\mu+{u \over 2}\right) \sum_i \left(X^{11}_i+2 X^{22}_i\right) \; .
\end{eqnarray}
The eigenstates are easily worked out~\cite{Arrachea, Anfossi1},  and read
\begin{equation}
    |\psi (N_s,N_d)> = \mathcal{N} (\eta^\dagger)^{N_d} \tilde X^{10}_{k_1}
    \cdots \tilde X^{10}_{k_{N_s}}\ket{\mbox{vac}} \, ; \label{psi}
\end{equation}
Here $\mathcal{N}=\left[(L-N_s-N_d)!/(L-N_s)!N_d!\right]^{1/2} $ is a normalization factor; $\tilde X^{10}_k$ is the Fourier transform of the Hubbard projection operator $X^{10}_j$, i.e., $\tilde X^{10}_k= \sum_j {1\over \sqrt{L}}\exp(i { \pi\over L} j k) X^{10}_{j}$. Moreover, $\eta^\dagger =\sum_{i=1}^L X^{20}_i$ is also known as the eta operator, commuting with $H$; $(\eta^\dagger)^{N_d}$ creates $N_d$ pairs which are fully spread over the chain. These are the $\eta$ pairs first introduced by Yang~\cite{ODLRO}. This structure corresponds to a very simple physical picture: eigenstates contain $N_s$ spinless fermions in momentum eigenstates $\{ {2 \pi k_1 \over L} , \dots, {2 \pi k_{N_s} \over L} \}$ and $N_d$ spinless bosons ($\eta$ pairs). \\
The energy eigenvalues are given by $ E(\{n_k\}, N_d)= -2 \sum_{k=1}^{L}  \cos( { 2 \pi k \over L } ) n_k - 2\mu N_d- (\mu + \frac{u}{2} ) N_s  $  where $n_k={0,1}$ is the number of fermions with momentum $ { 2 \pi k \over L }$. For any given $N_s = \sum_k n_k$ and $N_d$ the minimum is achieved by occupying with
$N_s$ fermionic particles the momentum modes $ \{ −\pi(N_s-1)/L, \dots, \pi (N_s-1)/L \}  $, the corresponding eigenvalue
being $ E(N_s,N_d)= -2  \sin \left(\pi\frac{N_s}{L}\right)/\sin \left(\frac{\pi}{L}\right)-2\mu N_d- \left(\mu + \frac{u}{2}\right)N_s \, $, whence we obtain the ground state energy density in the TDL
\begin{equation}
    \mathcal{E}(n_s, n_d)= - \frac{2}{\pi} \sin (\pi n_s) -2\mu n_d- \left(\mu + \frac{u}{2}\right) n_s \label{Energy}
\end{equation}
with $ \mathcal{E}=E/L, n_s=N_s/L, n_d = N_d/L$.  \\
The actual ground state is found by requiring that $n_s$ and $n_d$ minimize (\ref{Energy}). \\
For $\mu < 0 $ we have $n_d = 0 $, hence upon minimizing we get $n_s = \frac{1}{\pi} \arccos\left( -\frac{\mu}{2} - \frac{u}{4}\right) $. For $ -4 - 2 \mu \leq u \leq 4 - 2 \mu  $ we have empty and singly occupied sites (phase I), for $u > 4 - 2 \mu $ we have only singly occupied sites (phase IV) and for $ u < -4 -2 \mu  $ we have only empty sites. \\
For $\mu > 0 $ we have $n_d = (1-n_s)$, hence upon minimizing we get $n_s = \frac{1}{\pi} \arccos\left(\frac{\mu}{2} - \frac{u}{4}\right) $. For $ -4 + 2 \mu \leq u \leq 4 + 2 \mu  $ we have doubly and singly occupied sites (phase I'), for $u > 4 - 2 \mu $ we have only singly occupied sites (phase IV) and for $ u < -4 -2 \mu  $ we have only doubly occupied sites. \\
For $\mu=0$ we get $n_s = \frac{1}{\pi} \arccos\left( -\frac{u}{4} \right) $.  For $ -4 \leq u \leq  4 $ we have empty, doubly and singly occupied sites (phase II), while for $u > 4 - 2 \mu $ we have only singly occupied sites (phase IV) and for $ u < -4 -2 \mu  $ we have empty and doubly occupied sites (phase III). \\
Hence we get in the $\mu$-$u$ plane the phase diagram depicted in Fig. \ref{FGS1}, right. In the left part, the same ground-state phase diagram is drawn in the $n$-$u$ plane (with $n=N/L$ average per-site filling). The phase diagram presents various QPTs driven by parameters $u$ and $\mu$ (or $n$).
Each transition is characterized by a change in the number of on-site levels involved in the state. Phase IV has just one level per site since each site is singly occupied.
Phases I and I' (which is the particle-hole counterpart of phase I) have two on-site levels: singly occupied sites and empty or doubly occupied sites respectively. This holds for phase III as well, where only empty and doubly occupied sites appear. Phase II is the only phase in which all three on-site levels are involved.
Phases II and III are characterized by the occurrence of off-diagonal long-range order (ODLRO) and superconducting correlations, evaluated as:
 \begin{equation}\label{ODLRO}
     \lim_{r\rightarrow\infty}\langle X^{20}_i  X^{02}_{i+r}\rangle = n_d(1-n_d-n_s)\, .
\end{equation}
 Note that ODLRO ---though not allowing real superconducting order at $x=1$ due to spin degeneracy, which implies the vanishing of spin gap, is at the very root of superconducting order, which occurs at $x\neq 1$.~\cite{Anfossi4}\\
Before discussing the various transitions in terms of the QD behavior, let us recall some feature of each of them in terms of standard theory. First of all, since $N_d$ and $N_s$ are both conserved quantities, the transitions should be originated from level crossing. Indeed, they also occur at finite $L$. Nevertheless, none of them is of first order, since it can be easily checked that the first derivative of $E_{GS}$ is always smooth. In fact, the transitions I (I') $\to$ IV and II $\to$ IV and II $\to$ III are second-order QPTs, while the transition II $\to$ I (I') is an infinite-order QPT.

\subsection{Reduced density matrices}

The present work focuses on two-point correlations. To evaluate them, knowledge of the ground-state reduced density matrices is necessary, and we shall report their expression for completess (for a full derivation, the reader may refer to Refs. \cite{Anfossi2,Giordak}).  Correlations can be analyzed within two different and complementary pictures. Obviously, one can examine correlations between sites of the lattice (direct lattice picture). In addition, the structure of eigenstates in the model suggests yet another approach, namely to consider the reciprocal lattice, whose elementary nodes are momentum modes $k_j = \frac{  2 \pi}{L}j$, ${j=0, \dots, L-1}$. In some respects, the reciprocal lattice picture affords a simpler description of the system~\cite{Anfossi1, Anfossi3}. \\
Let us start by giving reduced density matrices the direct lattice picture. The one-site reduced density matrix $\rho_i $ when expressed in terms of the basis $\{\ket{0},\ket{1},\ket{2}\}_i $ is diagonal in all the regions of the phase diagram:
\begin{equation}
    \rho_i=\mbox{diag } \{1-n_s-n_d,n_s,n_d\}\quad ,\label{rhoi}
\end{equation}
the two-site reduced density matrix $\rho_{ij}$ in the basis  $\{ \ket{00}, \ket{01}, \ket{02}, \ket{10}, \ket{11}, \ket{12}, \ket{20}, \ket{21}, \ket{22}  \}_{ij} $ reads~\cite{Anfossi2}:
\begin{equation}
\rho_{ij}= \left( \begin{array}{ccccccccc}
D_1  & 0      & 0    & 0    & 0    & 0     & 0  & 0   & 0    \\
0    & O_1    & 0    & O_2  & 0    & 0     & 0  & 0   & 0    \\
0    & 0      & Q    & 0    & 0    & 0     & Q  & 0   & 0    \\
0    & O_2^*  & 0    & O_1  & 0    & 0     & 0  & 0   & 0    \\
0    & 0      & 0    & 0    & D_2  & 0     & 0  & 0   & 0    \\
0    & 0      & 0    & 0    & 0    & P_1   & 0  & P_2 & 0    \\
0    & 0      & Q    & 0    & 0    & 0     & Q  & 0   & 0    \\
0    & 0      & 0    & 0    & 0    & P_2^* & 0  & P_1 & 0    \\
0    & 0      & 0    & 0    & 0    & 0     & 0  & 0   & D_3  \\
\end{array} \right)\, .\label{rhoij}
\end{equation}
Here
\[
    \begin{array}{cclccclcc}
        D_1 & = & P_{ij} (1-c)^2\, ,\quad & & \;  O_2 &=& \Gamma_{ij}(1-c)\, ,\\
        D_2 & = & n_s^2- |\Gamma_{ij}|^2  \, ,\quad & & \; P_1 &=&
        c\left( 1-n_s -P_{ij}\right )\, ,\\
        D_3 &=& c^2 P_{ij}\, ,\quad
        & & \; P_2 &=& c \Gamma_{ij}\, ,\\
        O_1 &=& \left(1-n_s-P_{ij}\right) (1-c)\, ,\quad & & \; Q &=&
        c(1-c) P_{ij}\, ,\\
    \end{array}  \label{entries}
\]
with $c=n_d/(1-n_s)$, $P_{ij}=(1-n_s)^2-|\Gamma_{ij}|^2$, and $ \Gamma_{ij}= \frac{\sin (n_s \pi |i-j|)}{ \pi |i-j|)}\, . $ \\
Let us now turn to the reciprocal lattice picture. To each momentum mode $k_j$ corresponds a 4-dimensional Hilbert space, spanned by the basis
\begin{equation}
\mathcal{B}_{k_j}=\ket{0}_{k_j}, \ \ket{\uparrow}_{k_j}, \  \ket{\downarrow}_{k_j}, \  \ket{\uparrow \downarrow}_{k_j}, \     \label{recbasis}
\end{equation}
The reduced density matrix for any such mode reads, in the TDL, and in the basis (\ref{recbasis}),
\begin{equation}
\rho_{k_j} = \mbox{diag} (a^2, ab, ab, b^2)   \label{recstate}
\end{equation}
where $a = \frac{1-n_s-n_d}{1 - n_s}$ and $b = \frac{n_d}{1 - n_s} $. \\
The two-mode ($16 \times 16$) reduced density matrix for modes $k_i$ and $k_j$, $k_i \neq k_j$, is diagonal with respect to
the local basis $\mathcal{B}_{k_i} \otimes \mathcal{B}_{k_j}$. In the TDL, the eigenvalues are $a^{\alpha} b^{4-\alpha}$ with multiplicity $ m_{\alpha} = {4 \choose \alpha } $. \\
The case $k_i = -k_j$ has to be treated separately. The two-mode ($16 \times 16$) reduced density matrix for modes $k_j$ and $-k_j$ has support on a $4 \times 4$ subblock. Indeed the sole states that can be built by the action of the $\eta_{k_j}^\dag$ operators belong to the subspace spanned by: \begin{equation}
\mathcal{B}_{k_j , -k_j}= \{ \ket{0 , 0}_j, \ \ket{\uparrow , \downarrow}_j, \ \ket{\downarrow , \uparrow}_j, \
\ket{\uparrow \downarrow , \uparrow \downarrow}_j   \}   \label{recbasis2}
\end{equation}
where $\ket{\alpha \ \beta}_{j} \equiv \ket{\alpha}_{k_j} \otimes \ket{\beta}_{-k_j}$. \\
In the TDL, and in this basis, the nonvanishing subblock of the matrix reads:
\begin{equation} \label{recmatrix}
\rho|_{\mathcal{B}_{k_j, -k_j}} = \left( \begin{array}{cccc} a^2 & 0 & 0 & 0 \\
0 & ab & ab & 0 \\
0 & ab & ab & 0 \\
0 & 0 & 0 & b^2 \\
\end{array} \right).
\end{equation}

\subsection{Behavior of entanglement at QPTs}

Two-point entanglement at the QPTs of the model was thoroughly analysed in Refs.~\cite{Anfossi1,Anfossi2,Anfossi3,Giordak}, upon consideration of different correlation measures: the two-point concurrence $K_{i,j}$ or the two-point negativity $\mathcal{N}_{i,j}$ as measures of entanglement (notice that definition of concurrence is available for two-qutrit systems), the mutual information $I_{i,j}$ as a measure of total two-point correlations, and the single site entropy $\mathcal{S}_i$ as a measure of multipartite entanglement between one site and the rest of the chain. The behavior of all correlation measures was studied as a function of $x$ ($x=\mu$ or $x=u$) in the vicinity of the quantum critical points. Results are briefly summarized in the table below.

\begin{center}
\begin{tabular}{|c|c|c|c|c|c|c|} \hline
transition & x & $\frac{d\mathcal{S}_i}{dx}$ & $\frac{dI_{i,j}}{dx}$ & $\frac{dK_{i,j}}{dx}$ & $\frac{d\mathcal{N}_{i,j}}{dx}$ & ent \\ \hline
I $\to$ IV  &  $\mu$  & $\propto \frac{1}{\sqrt{\mu -\mu_c}}$ &$ \propto \frac{1}{\sqrt{\mu -\mu_c}}$ & $\propto \frac{1}{\sqrt{\mu_c -\mu}}$ & & Q2\\ \hline
II $\to$ I  & u &  $ \propto \log(u_c-u)$ & finite  &  & finite & QS \\ \hline
II $\to$ III  & u  & $\propto \frac{1}{\sqrt{u -u_c}}$  &  $\propto \frac{1}{\sqrt{u-u_c}}$ &  & finite & Q2\\ \hline
II $\to$ IV   & u & $\propto \frac{1}{\sqrt{u_c -u}}$  & $\propto \frac{1}{\sqrt{u_c -u}}$ & & finite & Q2 \\ \hline
\end{tabular}\label{Table: QSvsQ2}
\end{center}

The analysis of divergences allows to classify the different transitions into those driven by two-point correlations (Q2: II $\to$ III, II $\to$ IV, I $\to$ II), where some two-point correlation measure ($K_{i,j}$, $\mathcal{N}_{i,j}$ or $I_{i,j}$) diverges, and those driven by multipartite correlations (QS: II $\to$ I) where only $S_i$ diverges. However, the two-point charachter of the transitions II $\to$ III, II $\to$ IV is only detected by $I_{i,j}$ (a measure of total correlations), while $N_{i,j}$ (the measure of quantum correlations used) is unfit to discriminate between those transitions and the multipartite-driven one (II $\to$ I). 


\section{Results}
\label{Sec.: results}

\subsection{Region I (I')}
\label{Sec.: regionI}

\begin{figure}[htbp]
\begin{center}
\includegraphics[width=0.22\textwidth]{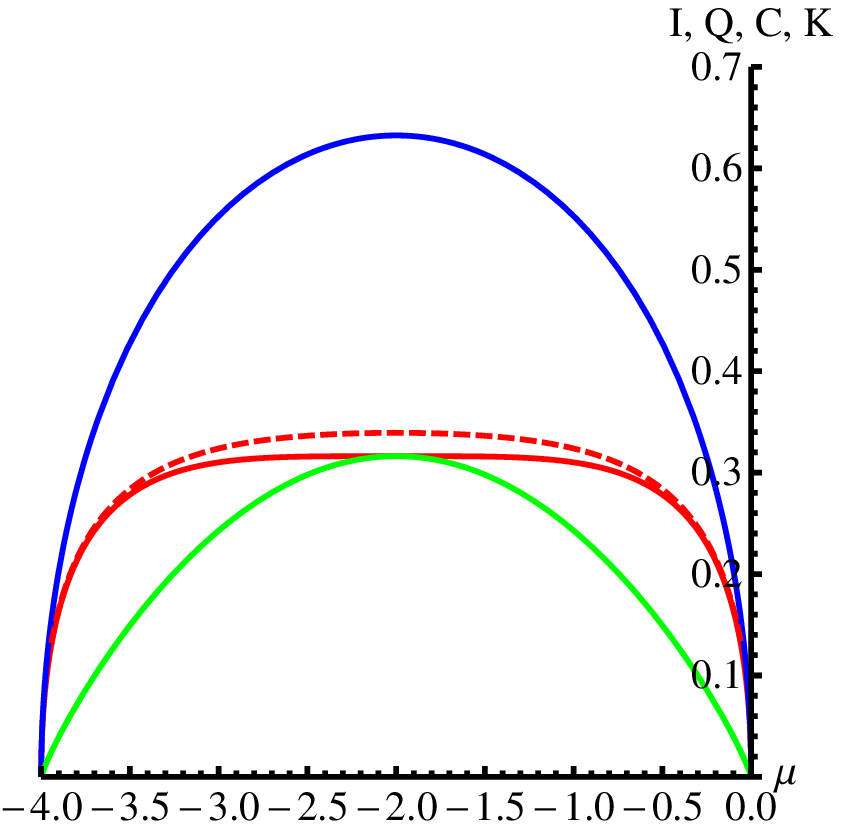}
\includegraphics[width=0.22\textwidth]{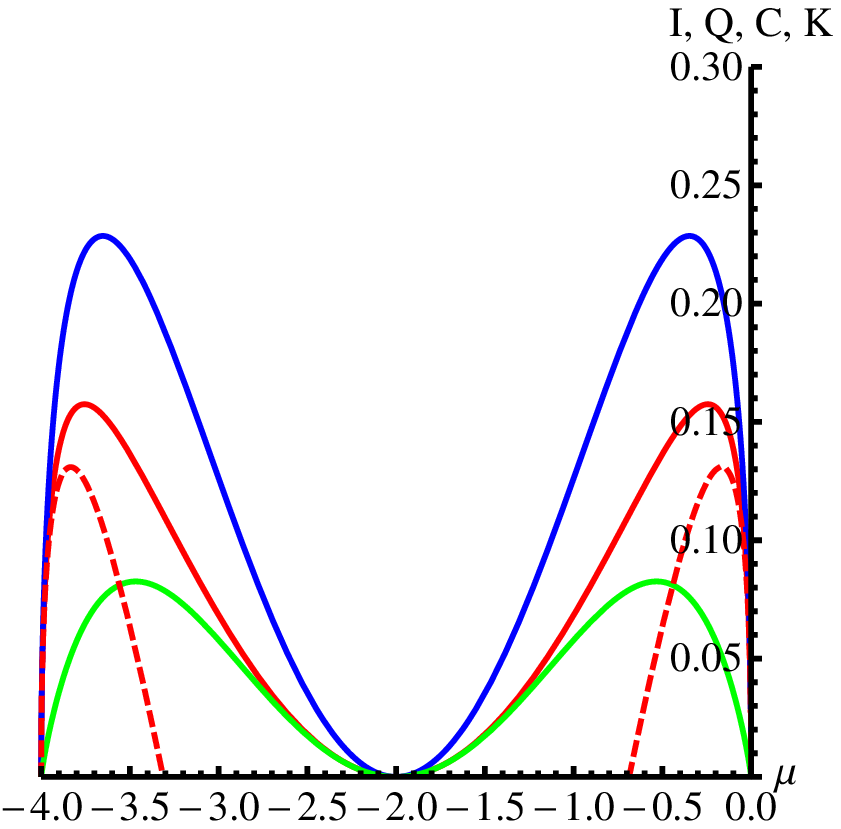}
\includegraphics[width=0.22\textwidth]{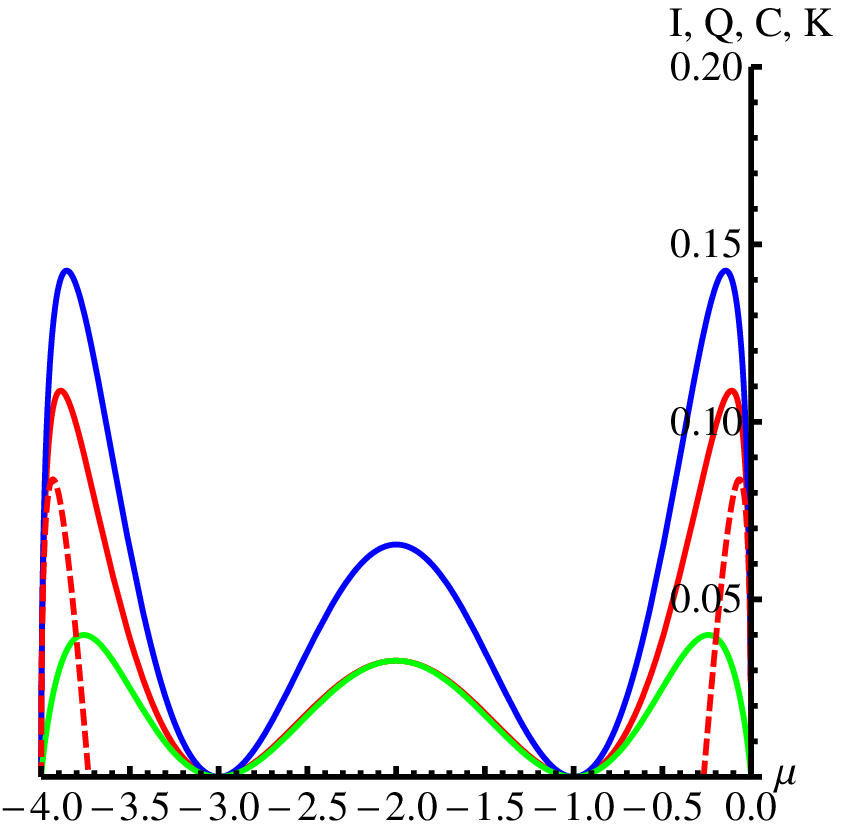}
\includegraphics[width=0.22\textwidth]{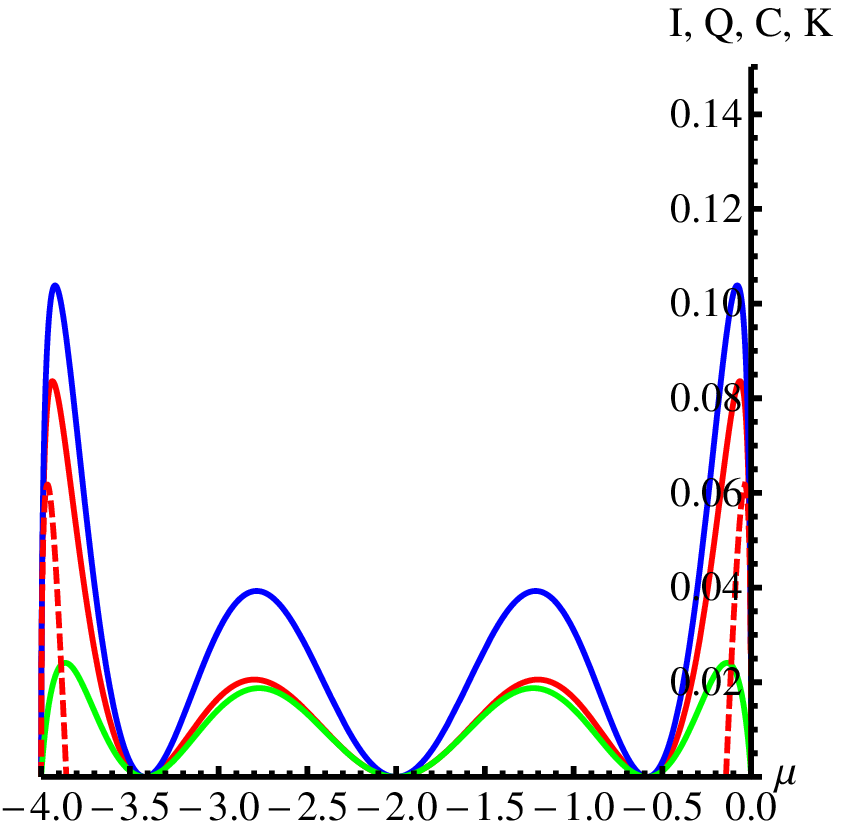}
\caption{ quantum mutual information $I_{i,j}$ (blue, solid), quantum discord $Q_{i,j}$ (red, solid), classical correlation $C_{i,j}$ (green, solid),  concurrence $K_{i,j}$ (red, dashed) as a function of $\mu$ in region I ($u=4$), for $|i-j|=1$ (top, left), $|i-j|=2$ (top, right), $|i-j|=3$ (bottom, left), $|i-j|=4$ (bottom, right).} \label{discord1}
\end{center}
\end{figure}

\begin{figure}[htbp]
\begin{center}
\includegraphics[width=0.4\textwidth]{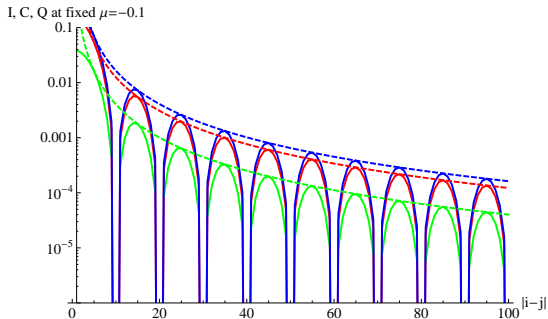}
\caption{ quantum mutual information $I_{i,j}$ (blue, solid), quantum discord $Q_{i,j}$ (red, solid), classical correlation $C_{i,j}$ (green, solid) as a function of $|i-j|$ in region I for $\mu=-0.1, u = 4$.
Upper dashed lines represent the envelope of the respective maxima wich exhibits a power law decay ($\sim |i-j|^{-2} $)
} \label{decay}
\end{center}
\end{figure}

We start our analysis by evaluating correlations (QD and CC) in phase I. Results for phase I' are omitted, since they are exactly equal (by virtue of the particle-hole symmetry one just has to replace emply with doubly occupied sites). \\ Phase I (I') is characterized by the absence of doubly occupied (empty) sites, so that the effective number of on-site levels reduces to $2$. Consequently, the 2-site reduced $9 \times 9$ density matrix $\rho_{ij}$ has nonzero entries only in the $4 \times 4$ subblock spanned by $\{ \ket{00}, \ket{01}, \ket{10}, \ket{11} \}_{ij} $ . \\
The quantum discord can be evaluated analytically through the methods developed in Refs~\cite{Luo, Ali, QChen,Girolami} (details are worked out in the Appendix, see \ref{qubit}). Evaluating the mimimum of the reduced conditional entropy reduces, Eq. (\ref{minimumqubit}), to taking the minimum among two functions, i.e., $\mbox{inf}_{\{B_k\}} S(\rho_{ij} | \{B_k\}) = \min\{ S_1, S_2 \}$,  where $S_1, S_2$ depend on $\theta_1 = \sqrt{(1+4n^2 - 4 n) + 4 |\Gamma_{ij}|^2}$ , $\theta_2 = \frac{|n - 2 n^2 + 2 |\Gamma_{ij}|^2 | }{n} $ , $\theta_3 = \frac{|1 + 2 n^2 - 3 n - 2 |\Gamma_{ij}|^2 | }{1 -n} $ (\ref{S1}-\ref{S3}). \\
We verify that for all values of $|i-j|$ we always have $S_1 \leq S_2 $ and therefore two-point classical correlation (\ref{Eq.: DefClassicalCorrs}) and quantum discord (\ref{Eq.: DefDiscord}) can be written in terms of $S_1$. \\ In order to compare quantum discord and entanglement, we also evaluate two-point concurrence~\cite{Anfossi2}
\begin{small}
\begin{equation}
K_{i,j} =  \min \left\{ 0, \left|\Gamma_{ij} - \sqrt{((1-n)^2 - |\Gamma_{ij}|^2) (n^2 -|\Gamma_{ij}|^2)} \right|  \right\}.
\end{equation}
\end{small}
In the following, letters $Q_{i,j}, C_{i,j}, I_{i,j}, K_{i,j}$ alway denote quantum discord, classical correlations, mutual information and concurrence respectively.
The values of $I_{i,j}$, $C_{i,j}$, $Q_{i,j}$ and $K_{i,j}$ for region I and different values of $|i-j|$ are plotted in Fig. \ref{discord1}. \\
\ \\
We first see that the $Q_{i,j}$ and $C_{i,j}$ have the typical oscillating behavior already shown by the mutual information \cite{Anfossi2}.
At variance with the previous analysis, where the quantum correlations measured by the concurrence were different from zero only in proximity of the borders of the regions i.e., for $\mu \rightarrow -4,0$, here we see that the system exhibits non zero discord within the whole region I except at some nodal points defined by
the equation $\Gamma_{ij} = \langle c^\dag_i c_j \rangle =  \frac{\sin (n_s \pi |i-j|)}{ \pi |i-j|)}= 0 $ where all correlation measures vanish, $I_{i,j}=C_{i,j}=Q_{i,j}=0$. Classical correlations show a similar behavior. Therefore, in the central region of phase I, where $K_{i,j}$ vanishes $ \forall |i-j|> 1$, two-point discord and classical correlations are still present.
Correlations are modulated by the sinusoidal behavior induced by $\Gamma_{ij}$ and {\it at fixed $\mu$} they all decay algebraically with the distance: $I_{i,j},Q_{i,j},C_{i,j} \simeq |i-j|^{-2}$, see Fig.\ref{decay}.

In proximity of the transition I $\to$ IV it was shown in \cite{Anfossi2} that the system exhibits an {\em entanglement transition}\cite{Amico}:  the range of the entanglement $\mathcal{R}_\mathcal{K}$, i.e., the maximal distance $|i-j|$ for which $K_{i,j} \neq 0$, goes to infinity when approaching the transition.
In particular, $K_{i,j}$ have a maximum value for $n_s \to 1$ as $|i-j| \to \infty$. This behavior is reflected in that of $I_{i,j},Q_{i,j},C_{i,j}$, which also exhibit a global maximum at a value $n_s^{(i,j)} \approx 1-1/(2|i-j|)$ which approaches $n_s = 1$ for $|i-j| \to \infty$. Hence, the behavior of discord mirrors that of the entanglement. This behavior is depicted in Fig.\ref{maxima}.
In fact, also the mutual information and the classical correlations exhibit the same kind of behavior. However the values of the maxima for the various measures $I_{i,j},Q_{i,j},C_{i,j}$ scale in a different way with the distance:
\begin{figure}[htbp]
\begin{center}
\includegraphics[width=0.3\textwidth]{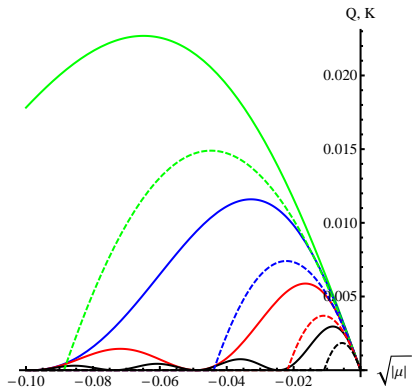}
\caption{maxima of quantum discord $Q_{i,j}$ (solid lines) and concurrence $K_{i,j}$ (dashed lines) for $ |i-j|=16 $ (blue), $|i-j|=32$ (red), $|i-j|=64$ (green), $|i-j|=128$ (black) as a function of $\sqrt{|\mu|}$ in region I ($u=4$) } \label{maxima}
\end{center}
\end{figure}

\bae
 I(n_s^{(i,j)},|i-j|)&\simeq&  \frac{1}{|i-j|} \\
 Q(n_s^{(i,j)},|i-j|)&\simeq&  \frac{1}{|i-j|}  \\
C(n_s^{(i,j)},|i-j|)&\simeq&  \frac{\log(|i-j|)}{|i-j|^2} \\
\eae
Therefore, when approaching the metal insulator transition I$\to$IV the  maxima of correlation measures ($I_{i,j}, Q_{i,j}, C_{i,j}$) decay in algebraic way along the chain. Quantum discord dominates for high distances, since the spreading of the classical correlation is suppressed by a factor $\log(|i-j|)/|i-j|$. This difference in the behavior of QD and CC defines the different role that they have at the transition and can be further appreciated by studying the derivatives of the different correlation measures with respect to $\mu$. In the critical limit $\mu \to 0,-4  $  we have
\be
\partial_{\mu} Q(\rho_{ij}) \simeq - \frac{1}{\pi \sqrt{|\mu- \mu_c|}}
\ee
while
\be
\partial_{\mu} C(\rho_{ij})  \simeq \frac{1}{\pi^2} \log|\mu - \mu_c|
\ee
Therefore, while the $\partial_\mu Q_{i,j}$ correctly agrees with the scaling behavior of $\partial_\mu I_{i,j} $ and $\partial_\mu K_{i,j} $ evaluated in \cite{Anfossi2}, $\partial_\mu C_{i,j}$ though being singular has a lower degree of divergence, so that classical correlations are subleading in the vicinity of the critical point. \\
We therefore see that the introduction of the new measures of correlations $Q_{i,j}$ and $C_{i,j}$ and the study of their derivatives allows on one hand to properly identify the metal-insulator transition and to properly classify it as a
two-point QPT~\cite{Anfossi2}, and on the other handallows for a refinement in description of the QPT. The importance of this feature will be more evident in the following paragraphs where we will describe the other two-point QPTs i.e., II $\to$ IV and II$\to$ II.
We close this subsection by discussing the role of the divergences of the different correlation measures and their relation with the divergences of the energy density of the system.
In ~\cite{Wu1}, the authors found  a direct relationship between the singularities (discontinuities and divergences) in the derivatives of the energy density of the system $\mathcal{E}=E/L$ with respect to the parameter $\lambda$ that drives the QPTs, and the singularities in the elements of the two-point reduced density matrix $\rho_{ij}$ or their derivatives with respect to $\lambda$.
In our case the,  the divergences in $\partial_{\lambda} I_{i,j}$ and $\partial_{\lambda} Q_{i,j}$ inherit the non analyticities of the derivatives of the elements of $\rho_{ij}$  at the critical point. In particular the elements:
\be
\partial_\lambda D_2 , \partial_\lambda O_1, \partial_\lambda O_2 \to  \frac{1}{\sqrt{|\lambda-\lambda_c|}}
\ee
show the same divergences exhibited by the second derivative of the energy density (\ref{Energy}) with respect to $\lambda=\mu$ (I$\to$IV), i.e., $\partial_{\mu}^2 \mathcal{E}\sim 1/\sqrt{|\mu-\mu_c|}$.
However, as we have seen above, classical correlations, though diverging, show a logarithmic divergence instead of an algebraic one, and accordingly one might believe that this is an accidental fact due to the definition of the correlation measure (i.e., that CC always display a lower degree of divergence). However, as we will see in the next sections, the classical correlations $C_{i,j}$ behave like $Q_{i,j}$ and $I_{i,j}$, in terms of their derivatives with respect to $\lambda=u$, at the transition II$\to$III, and therefore they coherently behave as the energy density at that transition i.e., $\partial_{u}^2 \mathcal{E}\sim 1/\sqrt{|u-u_c|}$. In summary, while the derivatives of different elements of $\rho_{ij}$ and of some of the correlations measures defined on the $\rho_{ij}$ show the same divergent behavior at the various transitions, which agrees with that of the energy density $\partial_{\lambda}^2 \mathcal{E}$, the classical correlations may show different kind of divergences and are thus able to discriminate between quantum phase transitions that are physically different.

\subsection{Region III: discord and ODLRO}
\label{Sec.: regionIII}

\begin{figure}
\includegraphics[width=0.3\textwidth]{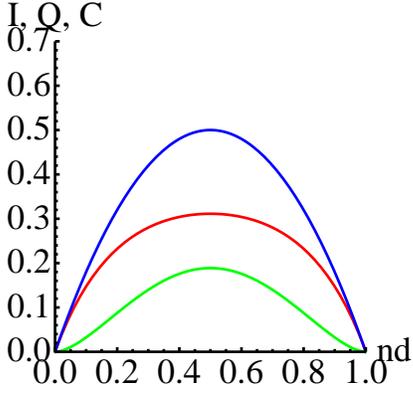}
\caption{ quantum mutual information $I$ (blue), quantum discord $Q$ (red), classical correlation $C$ (green)
as a function of $n_d$ in region III} \label{discordIII}
\end{figure}

Phase III is characterized by the absence of singly occupied sites, so that the number of on-site levels effectively reduces to $2$,
and the quantum discord can be evaluated analytically in the same way as above. Moreover in this case the number of Sutherland species reduces to 1. The quantum discord may be evaluated in the same way as above. We have $\mbox{inf}_{\{B_k\}} S(\rho_{ij} | \{B_k\}) = \min\{ S_1, S_2 \}$, where $S_1, S_2$ depend on
$\theta_1 =  (1-n_d)^2 +n_d^2$ and $S2$ on $\theta_2  =  1 - 2 n_d , \theta_3  =  1 - 2 n_d$ (\ref{S1}-\ref{S3}).
Since two-site density matrices $\rho_{ij}$ are equal for all $i,j$, the values of two-site correlations are equal for ech pair of sites, $I_{i,j}=I$,$C_{i,j}=C$,$Q_{i,j}=Q$. We have $S_1 \leq S_2 $ and therefore the classical correlations (\ref{Eq.: DefClassicalCorrs}) and the discord (\ref{Eq.: DefDiscord}) can be written in terms of $S_1$.\\
The values of $I$, $C$ and $Q$ for region III are plotted in Fig. \ref{discordIII}. \\
The first result of our analysis is that while in the TDL the concurrence $K_{i,j} =  \min \{ 0, - 2 n_d^2 (1-n_d) ^2  \} = 0$ vanishes everywhere in region III, the discord is always different from zero in the region; we thus have that the $\eta$-pair states display two-point quantum correlations, though not in the form of entangled correlations but rather in the form of QD.  Moreover, we notice that QD, as well as CC, between any two sites has the same value, irrespective of their distance: this reflects the way in which the $\eta$-paring mechanism spreads the correlations equally along the whole chain.
The $\eta$-pairing is also the ground for the appearance of ODLRO, which follow directly from (\ref{ODLRO}). It is intuitive to suppose that these superconducting correlations might be related to some kind of two-point quantum correlations, and indeed many authors have tried to find such a relation, see for example~\cite{Vedraleta,Fan}. While a relation with the entanglement properties in $k$ space was found in~\cite{Giordak} in the case of for $\eta$-pairs and BCS states, in direct space this relation could not be established in terms of the concurrence since the latter vanishes in the TDL~\cite{Anfossi2}.

While ODLRO in $\eta$-pair states cannot be related to two-point entanglement, our analysis allows instead to connect the ODLRO to the two-point quantum discord. Indeed we find that in the TDL $Q_{i,j}=Q_2^{TDL} , \forall i,j$ and we have
\bae
& Q_2^{TDL}  = f(x) = \frac{1}{\log 4} [4 x \ \mbox{arctanh}(1 - 2 x) + x \log 16 + \\ \nonumber
& \sqrt{1 - 4 x} \log(-1 - \frac{2}{-1 + \sqrt{1 - 4 x}}) + \log(\frac{1}{(x-1)^2}) + \log x]  \label{Eq.: DiscijvsODRLO}
\eae
where $f(x)$ a monotonically increasing function of $x=n_d (1-n_d)$, i.e., of the ODLRO. The above analysis allows to establish a {\em direct relation between a fundamental quantum property such as ODLRO and the presence of two-point (two-qubit) discord}. It therefore seems that the important two-point quantum correlations necessary in direct space for the appearance of the ODLRO are represented by the discord and not by the entanglement.\\
We finally note that the presence of the ODLRO in $\eta$-paris states is reflected also by the behavior of CC, which also are a monotonically increasing function of $n_d(1-n_d)$. The relation between CC and ODLRO will be important in the discussion of the transitions described in the next section.

\subsection{Region II}
\label{Sec.: regionII}

Region II contains empty as well as  singly and doubly occupied sites, so that there are $3$ on-site levels. This means that the evaluation of discord and classical correlations is more difficult than in the previous cases.
In order to evaluate $Q_{i,j}$ and $C_{i,j}$ we used the two numerical recipes described in \ref{qutrit}. The two methods show perfect agreement in the value of the discord throughout the whole region, and this is a first indication of their reliability. A further element of confidence in the methods used is the fact that $Q_{i,j}$ and $C_{i,j}$ must be continuous in the transitions II $ \to$ I , III (since all matrix elements of $\rho_{ij} $ are): when we approach the phase boundaries, the numerical limits of $Q_{i,j}$ and $C_{i,j}$ in region II coincide with the analytical values determined in region I and III.
\begin{figure}
 \includegraphics[width=0.22\textwidth]{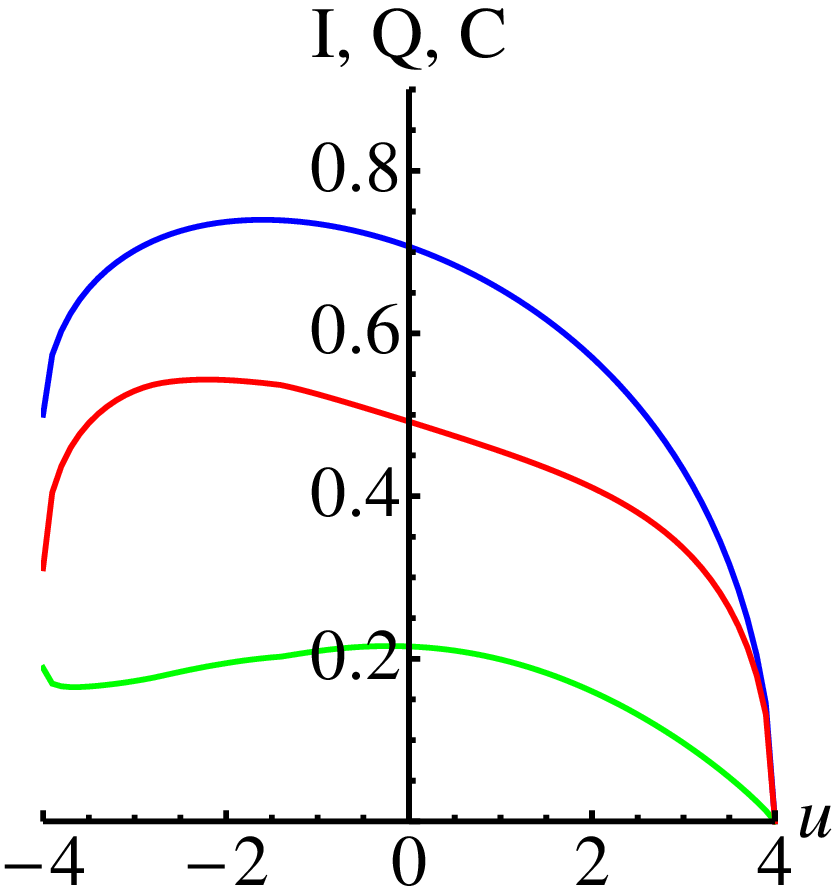}
 \includegraphics[width=0.22\textwidth]{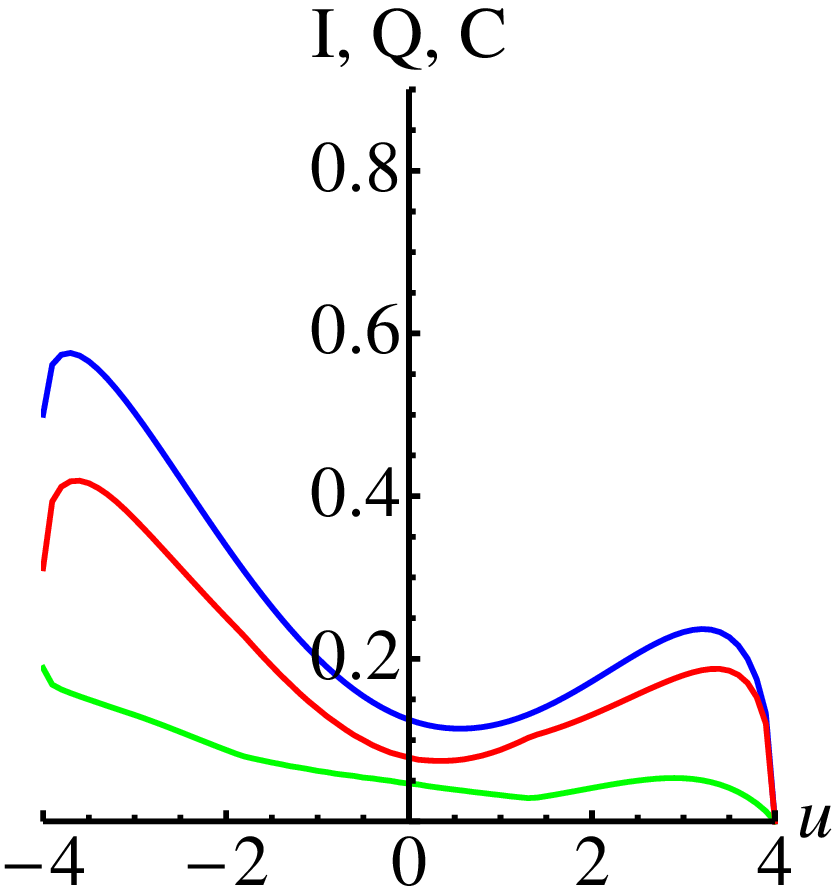}
 \includegraphics[width=0.22\textwidth]{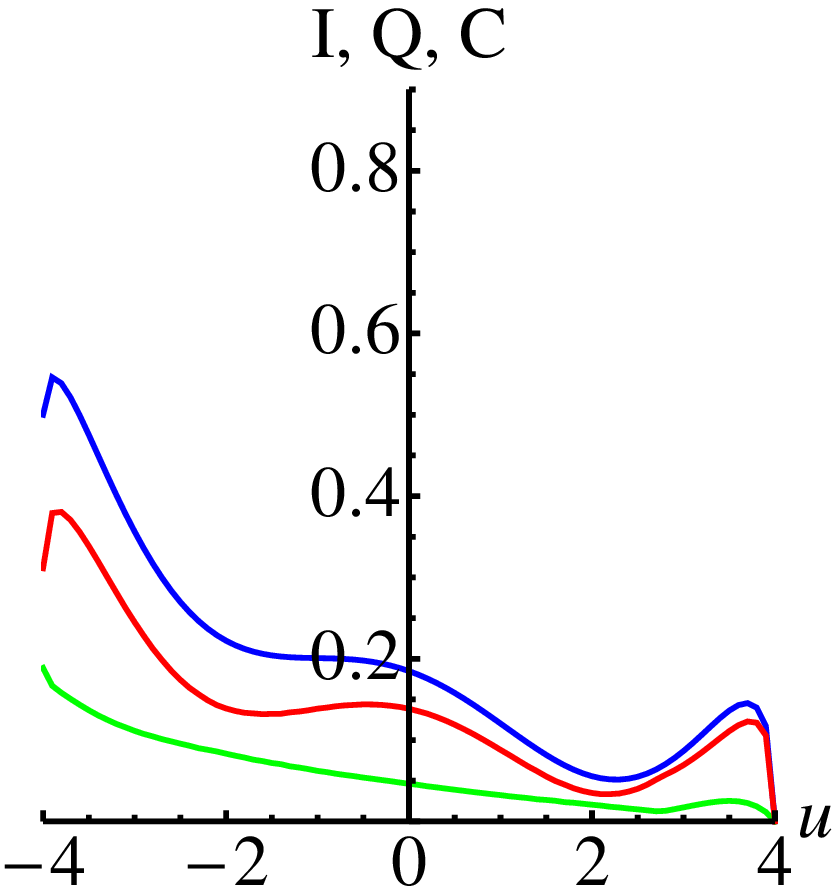}
 \includegraphics[width=0.22\textwidth]{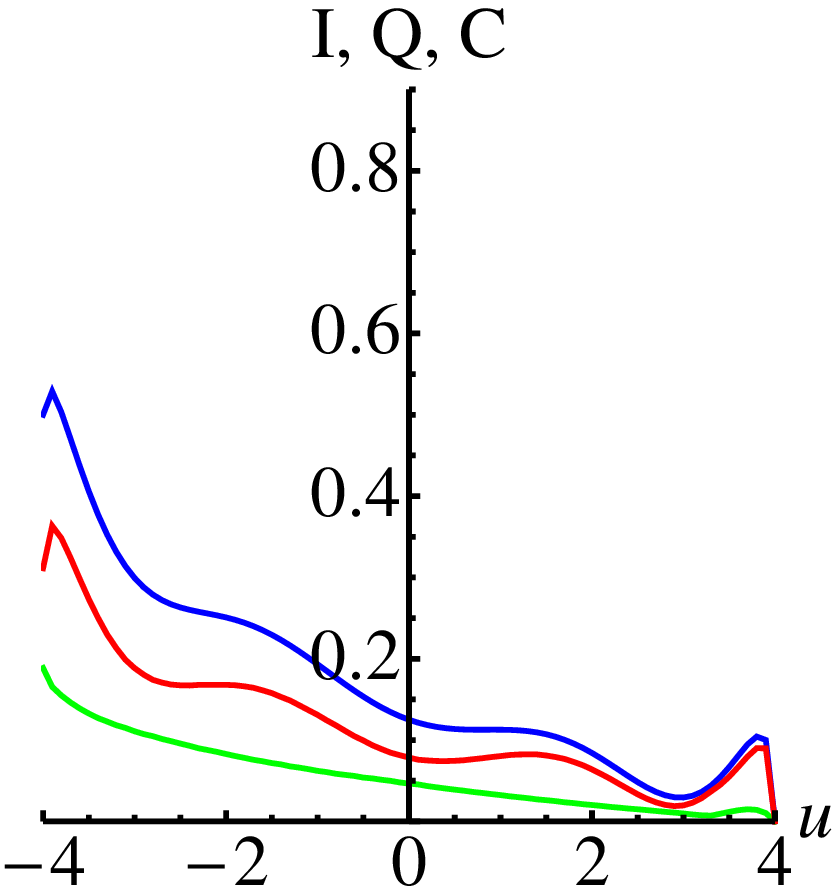}
\caption{ quantum mutual information $I_{i,j}$ (blue), quantum discord $Q_{i,j}$ (red), classical correlation $C_{i,j}$ (green)
as a function of $u$ in region II for $n=1$, $|i-j|=1$ (top, left), $|i-j|=2$ (top, right), $|i-j|=3$ (bottom, left), $|i-j|=4$ (bottom, right)} \label{discordII-e}
\end{figure}
\begin{figure}
 \includegraphics[width=0.22\textwidth]{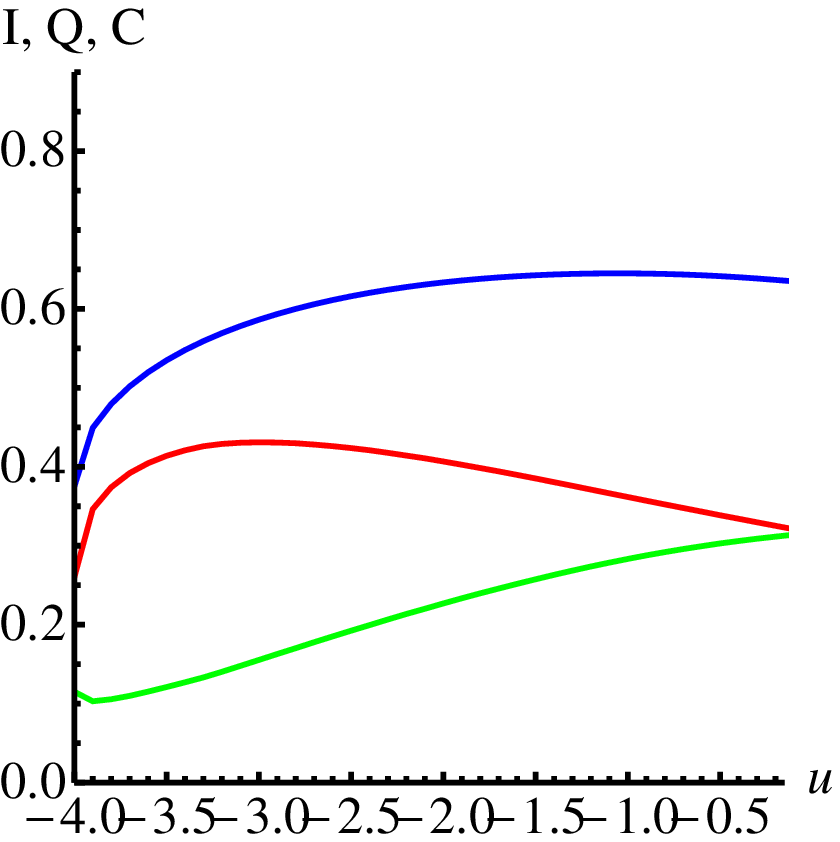}
 \includegraphics[width=0.22\textwidth]{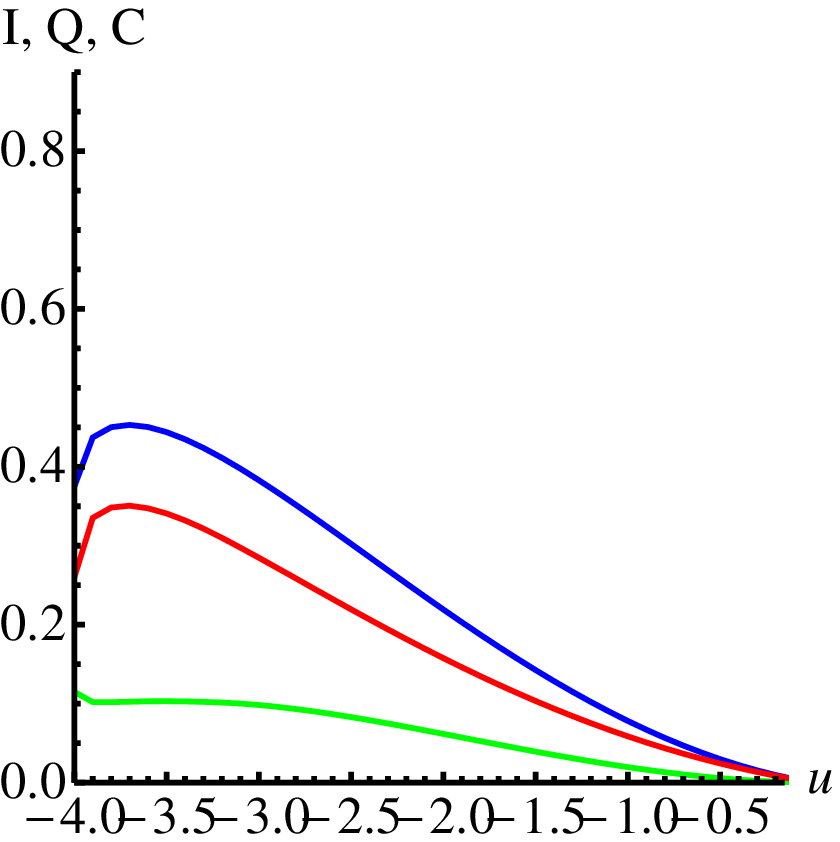}
 \includegraphics[width=0.22\textwidth]{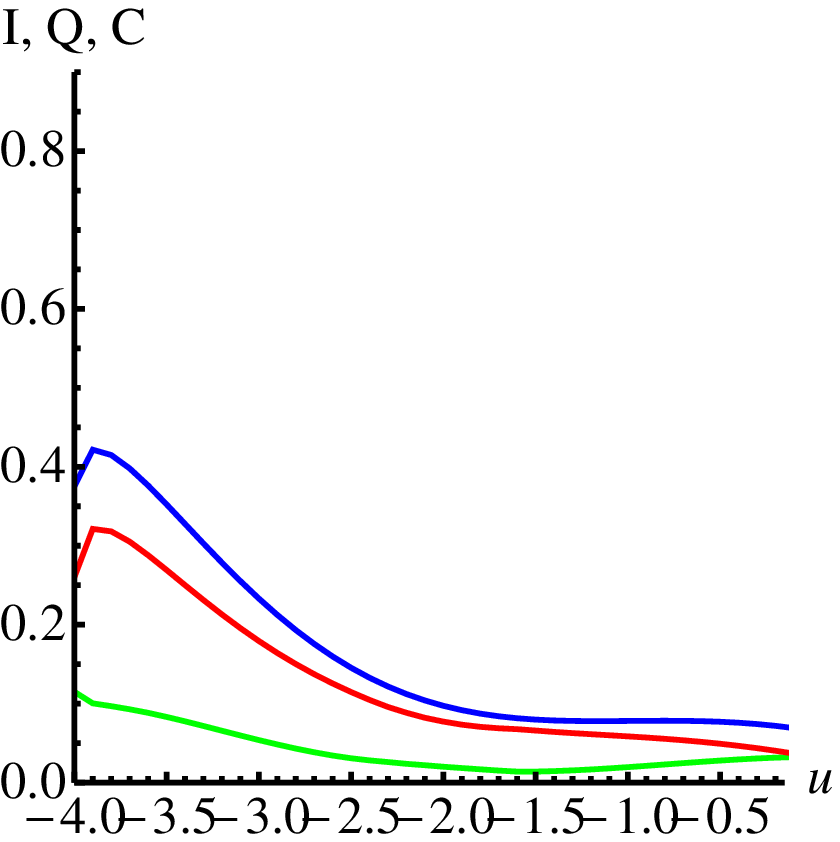}
 \includegraphics[width=0.22\textwidth]{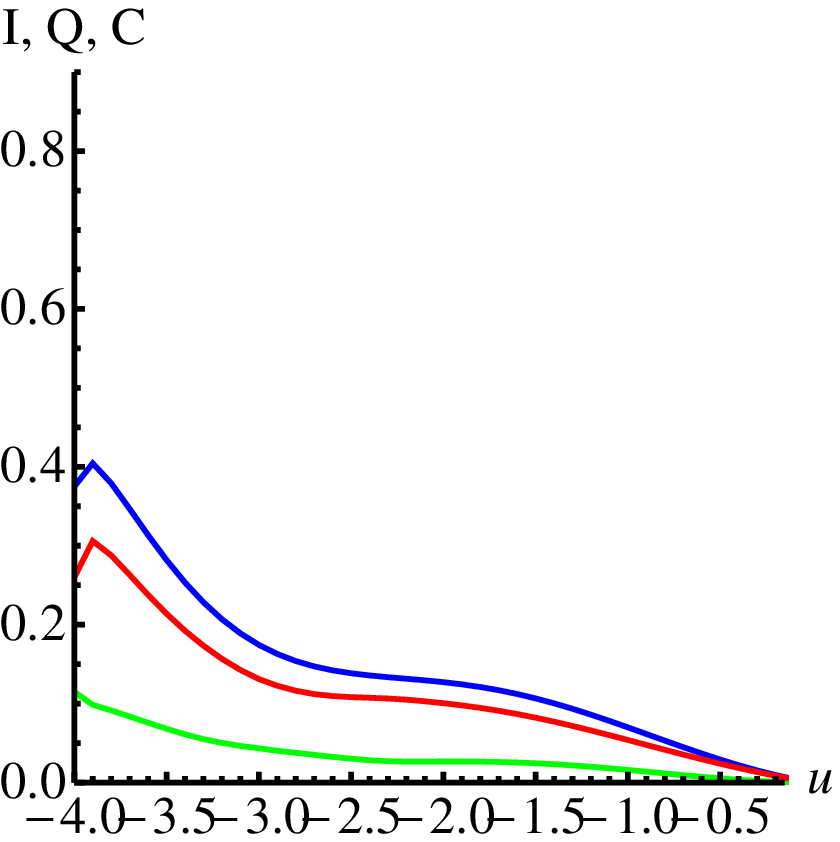}
\caption{  quantum mutual information $I_{i,j}$ (blue), quantum discord $Q_{i,j}$ (red), classical correlation $C_{i,j}$ (green)
as a function of $u$ in region II for $n=1$, $|i-j|=1$ (top, left), $|i-j|=2$ (top, right), $|i-j|=3$ (bottom, left), $|i-j|=4$ (bottom, right). We note that at the multipartite transition II$\to$I ($u=0$) all the two-point correlation measures behave in a smooth way} \label{discordII-f}
\end{figure}

In Fig. \ref{discordII-e} and Fig. \ref{discordII-f} we plot $I_{i,j}$, $C_{i,j}$, $Q_{i,j}$ in region II as a function of $u$ for $|i-j|=1,2,3,4$, and for $n=1$ and $n=0.5$ respectively. In the Table below, we summarize the critical behavior of the derivatives of quantum discord $Q_{i,j}$ and classical correlations $C_{i,j}$ for the transitions II $\to$ I, II $\to$ III, IV.  These  values are obtained as follows. We find numerically (with either of the procedures sketched above) the optimal measurement which minimizes the reduced conditional entropy. Contrary to what happens in region I, the orthogonal measurement minimizing the conditional entropy varies throughout region II, i.e., the parameters of the unitary rotation $V$ are not constant throughout the whole region. However, in the neighborhood of the critical lines ($ u \to -4$ and $u \to -4 \cos \pi n $) they are found to remain constant at any fixed $n$. We therefore use these constant values in the expressions for the reduced conditional entropy and obtain analytical formulas for $Q_{i,j}$ and $C_{i,j}$ as a function of $u$. We then extrapolate the critical behavior by studying these functions in the critical limit. \\
\ \\
\begin{center}
\begin{tabular}{|c|c|c|c|c|c|} \hline
transition & n &  u   & $\frac{dI}{du}$ & $\frac{dQ}{du}$ & $\frac{dC}{du}$ \\ \hline
II $\to$ I & $1/2$ & $\to 0$  & finite  & finite & finite \\ \hline
II $\to$ III & $1/2 $ & $\to -4$  & $\propto \frac{1}{\sqrt{u -u_c}}$  &  $\propto \frac{1}{\sqrt{u-u_c}}$ & $\propto \frac{1}{\sqrt{u_c -u}}$  \\ \hline
II $\to$ IV & $1 $ & $\to 4 $  & $\propto \frac{1}{\sqrt{u_c -u}}$  & $\propto \frac{1}{\sqrt{u_c -u}}$ & $\propto \log(u_c-u) $ \\ \hline
II $\to$ III & $1$ & $\to -4$  & $\propto \frac{1}{\sqrt{u -u_c}}$ &$ \propto \frac{1}{\sqrt{u -u_c}}$ & $\propto \frac{1}{\sqrt{u_c -u}}$ \\ \hline
\end{tabular}
\end{center}
\ \\
The results can be summarized as follows. In the transition II $ \to $ I o (or I') two-point $Q_{i,j}$, $I_{i,j}$, $C_{i,j}$ are regular, thus confirming that this transition has a multipartite nature.

As for the transitions II$\to$IV and II$\to$III, previous analyses \cite{Anfossi2} have shown that both transitions have a two-point character. As a first result, we see that at both transitions quantum discord is able to correctly detect the divergence expected, whereas negativity fails for this aspect\cite{Anfossi1} (see Sec. \ref{Table: QSvsQ2}).
The two transitions are however physically inequivalent, since they lead to two completely different phases: transition II$\to$IV is characterized by the disappearance of ODLRO, whereas at transition II$\to$III ODLRO is present.
We now show how this difference can be properly described by the study of the two-point classical correlations.\\
In the transition II $ \to $ III, while $\partial_u I_{i,j}, \partial_u Q_{i,j}>0$ and $\partial_u C_{i,j}<0$ all the derivatives display the same kind of algebraic singularity.
On the other hand, in the transition II $ \to $ IV, we have that $\partial_u I_{i,j},\partial_u Q_{i,j},\partial_u C_{i,j} <0$,  they all diverge, but $C_{i,j}$ has a lower degree of divergence
 i.e., it is logarithmic; this property allows to correctly describe the transition as a two-point one and furthermore to assimilate it to the metal-insulator transition I$\to$IV, where the CC show the same kind of divergence.\\
The result can be further  deepened by considering the following argument. All two-point correlations in region II can always be split into a finite and an infinite range contributions:  $A_{i,j}^{II} = \tilde A_{i,j} + A_{\infty}^{II} $, where $A=C,I,Q$ and $A_{\infty}^{II}=\lim_{|i-j|} A_{i,j}^{II}$. \\
The infinite range contributions can be analytically evaluated and they all explicitly depend  on the value of the ODLRO  in this phase, $n_d(1-n_s-n_d)$. Therefore, their derivatives with respect to $u$ have all the same behavior: they display the same type of algebraic singularity in case of transition II$\rightarrow$III (ODLRO), while they do not display any singularity in case of transition II$\rightarrow$IV (disappearance of ODLRO).\\
On the other hand, as for the finite range contributions we find that $\partial_u \tilde{C}_{i,j}$ diverges at both transitions but with a logarithmic behaviour that is dominant only in the transition  II$\rightarrow$IV (where $\partial_u \tilde{C}_{\infty}$ is regular) while its quantum counterpart $\partial_u \tilde{Q}_{i,j}$ diverges algebraically.
The above results show that the introduction of the discord and classical correlations allows to discriminate between two apparently similar but inequivalent two-point QPTs, and to root their difference in the persistence (disappearance) of ODLRO at the transitions.


\subsection{Reciprocal Lattice}
\label{Sec.: reclattice}

We now consider quantum discord between two momentum modes in the reciprocal lattice; the analysis is significant in region II and III, where $\eta$-pairs are present, and for values of $k_j>k_s $ where $ k_s = {2 \pi N_s \over L }$ is the maximum single-fermion momentum, since the portion of k-space pertaining to single fermions is factorized.
Let us fist consider two modes $k_j \neq k_j $. From the results derived in \cite{Anfossi3} we have that the measures of correlations all depend on a single parameter $a$ linked to the average occupation number of a generic mode $k_j$, $a=\average{n_{k_j}}/2=n_d/(1-n_s),\ \ \forall k_j$. In particular, the only pairs of modes $(k_i,k_j)$ which are correlated are the ones  for which $k_i=-k_j$, while if $k_i\neq-k_j$ the relative momentum modes are completely uncorrelated i.e., $I_{k_i, k_j } = 0 $ and therefore $Q_{k_i, k_j} = 0$. When $k_i=-k_j$ the single-mode von Neumann entropy reads $S(\rho_{k_j}) = -2 ( a \log a +  (1-a)\log (1-a))$, the two-mode von Neumann entropy is $S(\rho_{k_i, k_j})  = S(\rho_{k_j})+2 a(1-a)$ and hence the mutual information is $I_{k_i, k_j } = -2 (a \log a + b \log b - ab)$. \\
In order to evaluate the quantum discord, we should now consider the reduced conditional entropy after a generic measurement is performed on mode $k_j$, and minimize with respect to all measurements. It turns out that, a von-Neumann measurement $B= \{ \Pi_{0} , \Pi_{\uparrow},\Pi_{\downarrow}, \Pi_{\uparrow\downarrow} \} $ onto the trivial basis $\mathcal{B}_{-k_j}$ yields
\bae
\rho_{0} & =& \frac{1}{p_{0}} \mbox{Tr}_{-k_j} \Pi_{0} \rho_{k_j,-k_j} \Pi_{0} =  a^2 \ket{0} \bra{0} \nonumber \\
\rho_{\uparrow}  &=& \frac{1}{p_{\uparrow}} \mbox{Tr}_{-k_j} \Pi_{\uparrow} \rho_{k_j,-k_j} \Pi_{\uparrow} =ab \ket{\uparrow} \bra{\uparrow}\nonumber \\
\rho_{\downarrow} & =& \frac{1}{p_{\downarrow}} \mbox{Tr}_{-k_j} \Pi_{\downarrow} \rho_{k_j,-k_j} \Pi_{\downarrow} =ab \ket{\downarrow} \bra{\downarrow} \nonumber \\
\rho_{\uparrow \downarrow}  &=& \frac{1}{p_{\uparrow\downarrow}} \mbox{Tr}_{-k_j} \Pi_{\uparrow\downarrow} \rho_{k_j,-k_j} \Pi_{\uparrow\downarrow} =b^2 \ket{\uparrow\downarrow} \bra{\uparrow\downarrow}
\eae
so that $\sum_\alpha p_{\alpha} S(\rho_{\alpha}) = 0$ and the minimum is immediately attained.
Consequently we have that the quantum discord has a simple expression
\be
Q_{k_j,-k_j} = I_{k_j,-k_j}-S(\rho_{k_j})= 2 a(1-a)\propto \mathcal{N}_{k_j,-k_j} \label{remarkable}
\ee
and it is simply proportional to the negativity $\mathcal{N}_{k_j,-k_j}$\cite{Giordak}. .
This result allows us to derive some important conclusions. On one hand, the relationship found in \cite{Giordak} between ODLRO and negativity in region II can be rewritten in terms of the discord $Q_{k_j,-k_j}$ showing once again the quantum roots of the ODLRO:
\be
\lim_{|i-j|\to \infty} \langle X_i^{20} X_j^{02} \rangle = (1-n_s)^2 a(1-a)=(1-n_s)^2 Q_{k_j,-k_j}/2.
\label{Eq.: ODLRO}
\ee
This result, together with eq. \ref{Eq.: DiscijvsODRLO}, allows to establish a functional relation between the two-site discord $Q_{i,j}$ and the two mode discord $Q_{k_i,-k_i}$.

On the other hand, the line $n=1$ is an {\em iso-correlation} line \cite{Anfossi3}: since $a=1/2=const$, and therefore the momentum particle density $\average{n_{k_j}}$, and all the correlations between subsystem in the momentum picture are maximal and constant in the whole phase II. In particular, they are constant at the transition II$\to$III, therefore this transition cannot be identified by studying the derivatives of the correlation measures in $k$ space. On the other hand, at the transition II$\to$IV there is a sudden change in all correlations that discontinuously drop to zero in correspondence of the insulating phase that is characterized by a ground state which is factorized also in the momentum space i.e., $\ket{\psi}=\otimes_{k_j}\ket{\sigma}_{k_j}$, with $\sigma=\uparrow, \downarrow$. We therefore see that the two-point transitions II$\to$III and II$\to$IV can be distinguished even in momentum space, and this reinforces the result obtained in the previous section in the direct lattice picture, where the difference between the two transitions is highlighted by the behavior of $C_{i,j}$.

\subsection{Monogamy of quantum discord}
\label{Sec.: monogamy}

\begin{figure}
\includegraphics[width=0.24\textwidth]{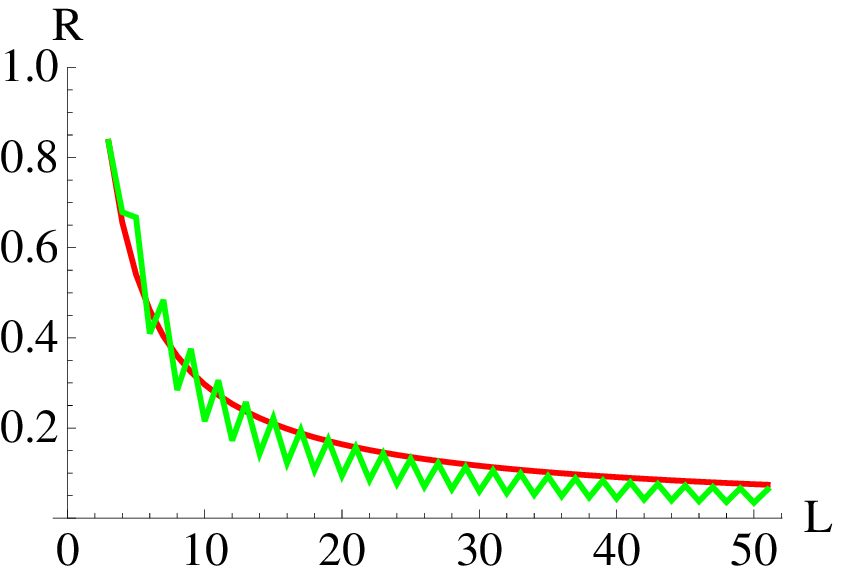}
\includegraphics[width=0.22\textwidth]{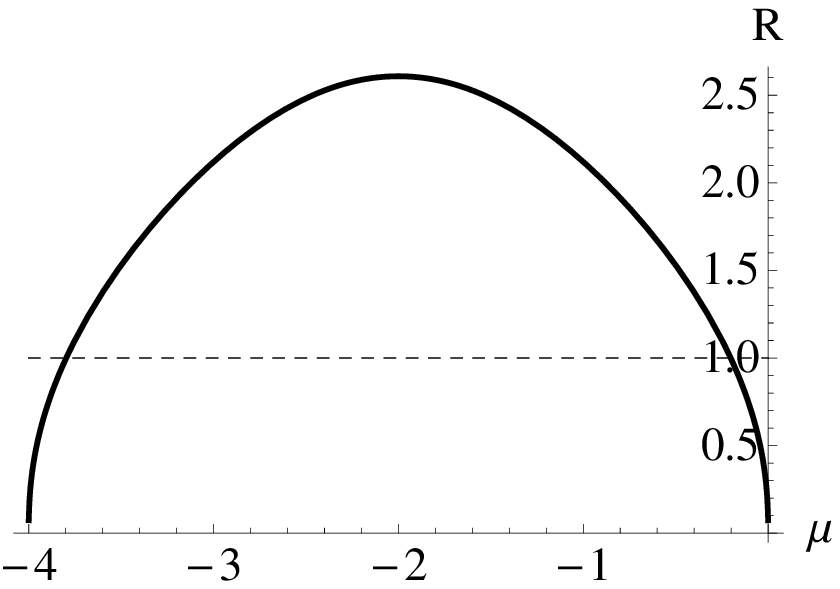}
\caption{Left: ratio $R$ in phase III for chains of varying length $L$, with $N_d=1/L$ (red), $N_d=\lfloor L/2\rfloor /L$ (green). Right: ratio $R$ in phase I in the TDL.}
\label{Fig.: poligamy}
\end{figure}

The study of the ground state properties of the extended Hubbard model can be fruitfully extended in order to assess a relevant quantum information problem: the relations between entanglement and discord. In this framework one of the main question to address is
the possibility that the discord may satisfy a  \textit{monogamy} relation similar to the one satisfied by the squared concurrence in the case of  $n$-partite qubit states~\cite{monogamy}:
\begin{equation} K_{1, (2 \dots L)}^2 \geq K_{12}^2 + \dots K_{1L}^2  \label{monogamyK} \end{equation}
where $K_{1, (2 \dots L)}$ is the concurrence between site $1$ and the rest of the system while $K_{1,j}$ is the concurrence between qubit $1$ and $j$.
Indeed, an analogous relation for the two-qubit discord has been recently discussed for the limited case of tripartite states case in \cite{monogamy_discord1} where it was found that the discord can be polygamous for tripartite $W$ states.
Here the discussion can be extended for the case of more general $n$-partite pure states, with $n\ge3$; in the following we will focus on two different case: the ground states in region III and I.

As for Region III, we have that the $\eta$-pairs states coincide with class of two-qubits permutational invariant states that can be written as
\be
\ket{\psi(N_d,L)}=\left ( \begin{array}{c}
                          L \\
                          N_d
                        \end{array}\right)^{-1}\sum_P P \ket{N_d,L-N_d}
                        \label{Permeta}
\ee
where $\left ( \begin{array}{c}
                          L \\
                          N_d
                        \end{array}\right)$ is the binomial coefficient, $\ket{N_d,L-N_d}$ is a fixed state with a given sequence of $N_d$ ones (pairs) and $L-N_d$ zeros (empty sites), and the sum is taken over all possible permutations $P$s (the $3$-partite $W$ state belongs to this class of states).
For states of these form, at fixed $N_d< L$ the single and two site reduced density matrices can be easily evaluated from (\ref{rhoi}) and (\ref{rhoij}) and they are equal for all sites, i.e., $\rho_i=\rho_1$ and $\rho_{ij}=\rho_2$ and the discord can be evaluated as described in the previous sections.
Since $\ket{\psi(N_d,L)}$ is a pure state the QD between one site and the rest of the chain is equal for all sites $Q_{(1|2,\cdots,L)}=Q_1$ and it simply coincides with the entanglement between the site and the rest of the chain, i.e., $Q_1=\mathcal{S}(\rho_1)$, it is a function of $n_d$ only and and it is bounded by $1$.
Both for finite  $N_d, L$ and in the TDL the two point QD $Q_{1,j}=Q_2(n_d,L)$ does not depend on $j$ and therefore $\sum_j Q_2 (n_d,L)= (L-1) Q_2(n_d,L)$.
As already mentioned, similar arguments can be applied to the concurrence $K$: with $n_s=0$ the dependence on $|i-j|$ disappears and, in particular for large $L$ one has $K_{1,j}\approx 1/L$;
for finite $N_d, L$ the concurrence is small but different from zero, and the monogamy property is always satisfied by the squared concurrence. \\
On the other hand, a direct evaluation of the above quantities shows that $\forall N_d \mbox{ and } L \ge 3$ one has $R=Q_1/[(L-1) Q_2(n_d,L)] < 1$; in Fig.~\ref{Fig.: poligamy} (left panel) we show the ratio $R$ for $N_d=1/L, \lfloor L/2\rfloor /L$ and different values of $L$.\\
While a general analytical demonstration of this result is not straightforward, one can note that in the case of permutational invariant states, for any fixed value of $n_d$ it is always possible to find an infinite number of states $\ket{\psi(N_d,L)}$ with $L=N_d/n_d$ and such that $Q_1\le (L-1) Q_2(n_d,L)$, i.e., the monogamy relation is violated. Indeed, while $Q_1$ just depends on $n_d$, $Q_2(N_d,L)$ is a decreasing function of $L$ which is lower bounded by its TDL expression (\ref{Eq.: DiscijvsODRLO}). Therefore all the states for which $L=N_d/n_d$ satisfies the relation $Q_1\le (L-1) Q_2^{TDL} $ will violate the monogamy relation.\\
As for the TDL, while $K_{1,j} \rightarrow 0$, $Q_s$ is constant at fixed $n_d$ and $Q_2=Q_2^{TDL}$ as in (\ref{Eq.: DiscijvsODRLO}) and therefore $R \rightarrow 0$.\\
Since the above arguments apply to a whole class of permutation invariant $n$-partite two-qubit states (\ref{Permeta}), we can state in full generality a property of two-qubit QD: \textit{ for $n$-partite states ($n \ge 3$) QD can be polygamous both in presence (for finite $N_d,L$) and in absence (TDL) of two point entanglement}.

While it is tempting to relate the violation of a monogamy relation by the discord to the presence of those correlations that are typical of $\eta$-pairs states, and that give rise in the TDL to ODLRO, our model shows that there are other classes of states in which such violation can occur. Indeed, in Fig.~\ref{Fig.: poligamy}(right panel) we report the ratio $R$ for the ground state of region I, which reads:
\be
\ket{\psi(N_s,L)}= \ket{k_1, \dots, k_{N_s}} = \tilde X^{10}_{k_1} \cdots \tilde X^{10}_{k_{N_s}} \ket{\mbox{vac}}
                        \label{fermiongs}
\ee
i.e., contains $N_s$ fermions in momentum eigenmodes ($k_1, \dots, k_{N_s}$), created by action of the Fourier transform of the Hubbard projection operator $\tilde X^{10}_k= \sum_j {1\over \sqrt{L}}\exp(i { \pi\over L} j k) X^{10}_{j}$ onto the vacuum.
The results refer to the TDL case and they show that for such states, although $Q_{i,j}$ does depend on the distance $|i-j|$, the monogamy property is violated by the two point discord in proximity of the QPT I $\rightarrow$ IV. This feature reflects the fact discussed in Section~\ref{Sec.: regionI} that when $\mu \rightarrow 0$, there is a spreading of the quantum correlations over the whole chain. Indeed, the violation of the monogamy condition starts in correspondence of $\mu\approx -0.2$, where the entanglement has already started to spread along the chain and has a finite range ($\mathcal{R}_N$ diverges only at the transition).\\
This result has two interesting consequences. On one hand the ground states in region I show that, depending on the parameters that define them ($n_s$ in this case) for the same class of states the discord may or may not violate a monogamy relation \cite{monogamy_discord1}.
On the other hand the behavior of the discord allows to refine the description about region I carried out in \cite{Anfossi2}. There, by means of the entanglement and correlation ratio it was pointed out that the ground states in region I have a truly multipartite character in the center of the region, while when approaching the transition the weight of the two-point correlations starts to increase; and this agrees with the two-point character of this transition.
Here this picture is revealed by the violation of the monogamy property displayed by the QD: in order to prepare the two-point transition at $\mu=0$, the system reorganizes its correlations in such a way that their two point character starts to prevail; one can therefore identify the point in which this process starts with the value of the parameters
i.e., $\mu\approx-0.2$ at which the monogamy property is violated by the discord.

We finally  compare the two above cases in terms of the violation of the monogamy property. Here the key observation is the different kind of violation exhibited by the discord. In region I the discord can be polygamous but the amount of quantum correlations shared by a single site with the other sites of the chain is finite i.e., $0<R<1$ for $\mu \neq 0$ and it vanishes at the transition $\mu=0$ because $Q_1\rightarrow 0$, while $\sum_j Q_{i,j}$ tends to a finite value . On the contrary, for $\eta$-pair states the violation has a completely different nature: each site can be equally correlated with all the other sites of the chain: $R\equiv0 \, \forall \, n_d$. This difference is indeed rooted in the presence of ODLRO in the TDL and in the previously found relation between discord and ODLRO. This kind  of violation is associated by the disappearance of the two-sites entanglement, while for the state in region I, the violation occurs in presence of bipartite entanglement.\\
The above results allows to give a general statement about quantum discord for multipartite pure states: it can be non-monogamous both in presence and in absence of bipartite entanglement. However the violation of the monogamy property can be maximal when ODRLO is established in the TDL and no bipartite entanglement is present in the state.

\section{Conclusions} \label{Sec.: conclusions}

In this paper we have addressed several important questions related to the ground state correlation properties of a reference fermionic model, the bond-charge Hubbard model. We have applied the recently developed measure of quantum discord (QD) and classical correlations (CC) to study how these relate to quantum phase transitions displayed by the model. By means of analytical and numerical analysis we have derived and analyzed the expressions of QD and CC for two-qubits and two-qutrits systems both in the direct lattice and in momentum space. Our results allow to describe the different quantum phase transitions in terms of the divergences of the various correlation measures. As shown in \cite{Anfossi1,Anfossi2, Anfossi3} the transitions can be  classified on the basis of the relevance of the two-point and multipartite correlations involved.
At variance with other entanglement measures \cite{Anfossi1}, such as negativity, QD (and CC) exhibits the expected non analyticities that define the two-point transitions.
Moreover, the comparison of their behaviour allows to discriminate between two apparently similar kind of two-point transitions.
In particular, a careful study of the contributions in which CC can be decomposed gives the possibility to detect the presence (disappearance) of the off diagonal long range order (ODLRO) and to identify its consequences at the various transitions.\\
Furthermore, the study of the discord between  two generic sites $i,j$ and two momenta modes $k_i,-k_i$ allows to establish a direct relation between ODLRO and the two-site/momenta modes discord, which turns out to be a monotonic function of ODLRO. This result is remarkable, since in the TDL no two-site entanglement is present in this states. By means of the same analysis it is possible to establish a functional relation  between the two-sites discord in direct space $Q_{i,j}$ and two-modes discord in momentum space $Q_{k_i,-k_i}$.

The study of $Q_{i,j}$ for $\eta$-pairs states is also important for describing the behavior of the discord with respect to the {\it monogamy} property. \cite{monogamy}. Indeed,  the $\eta$-pairs states are isomorphic to a relevant class of permutational invariant multipartite qubit states. While in the finite size case, all the states in the class display non-zero two-qubit entanglement, in the TDL the latter vanishes. However, in both cases we have shown that two-qubit discord is in general different from zero and furthermore it violates a monogamy relation.
Finally, we have shown for another class of states, the non-interacting fermionc ground states in region I, the discord can be polygamous depending on the values of the parameters. The main difference between the two class of states analyzed resides in the kind of violation of the monogamy property: only for the $\eta$-pair states the single qubit can be arbitrarily correlated with all the other infinite sites, thus leading to a maximal violation of the monogamy property. This fact is rooted in the presence of ODLRO in these states and in the direct relation between ODLRO and discord.

Our results confirm that the application of quantum information concepts to condensed matter systems can fruitfully lead to a precise description of the role of correlations in quantum phase transitions and at the same time to the development of useful relations that shed new light on the nature of quantum correlations as measured by discord.

\appendix

\section{Evaluation of discord}

The difficult step in evaluating the discord $Q$ is the minimization of the conditional entropy $ S(\rho_{ij} | \{B_k\} ) $ with respect the set of all von Neumann measurements. \\
In general, the concavity of the conditional entropy implies that its minimum is attained through extremal POVMs~\cite{Hamieh} and in particular through rank-1 POVMs, i.e., projective measurements~\cite{Dattaphd}.
Often it is sufficient to consider von Neumann measurements, but in some in some cases considering more general projective measurements allows for a better minimization, an issue still under research (see for instance Ref.~\cite{QChen}). \\
The minimization can be done analytically for some simple cases of two-qubits, namely for the class of X states which have nonnull entries only on the diagonal and antidiagonal and include states with maximally mixed marginals(see Refs.~\cite{Luo, Ali} and \cite{QChen,Girolami} for recent developments). On the contrary, the two-qutrit case must be handled numerically.

\subsection{two-qubit states} \label{qubit}
Since in phases I and III  the density matrix $\rho_{ij}$ corresponds to an X-state for which $\mbox{inf}_{\{B_k\}} S(\rho_{ij} | \{B_k\} ) $ can be easily evaluated with a fully analytical way by resorting to the method developed in ~\cite{Ali}. In this part of the Appendix we give a brief review of this method. \\
An arbitrary (single-qubit) von Neumann measurement is defined by a couple of orthogonal projectors $B_0$ and $B_1$, which can be obtained from $\ket{0} \bra{0}$ and $\ket{1} \bra{1}$ by an arbitrary $SU(2)$ rotation $V$:
\begin{equation}
B_0 = V \ket{0} \bra{0} V^\dag \qquad B_1 = V \ket{1} \bra{1} V^\dag
\end{equation}
Since $ V = tI + i \vec{y} \cdot \vec{\sigma} $ with $t^2 + y_1^2 + y_2^2 + y_3^2 = 1 $, von Neumann measurements are parametrized by three independent numbers. \\
The key result of ~\cite{Ali} is that the minimum of $ S(\rho_{ij} | \{B_k\} ) $ is always attained for some special values of the parameters $m = (ty_1+y_2y_3)^2$, $n = (ty_2-y_1y_3)(ty_1-y_2y_3)$,$ k = t^2 + y_3^2$,
namely
\bae
 &&\{k=0, m=0, n=0\} \mbox{and}  \nonumber \\
 &&\{k=1/2, m=0,1/4, n=0,\pm 1/8 \}
\eae
Therefore the minimization procedure reduces to comparing the expressions $ S(\rho_{ij} | \{B_k\} ) $ obtained in correspondence of these two sets of values. Furthermore,
when the two-site reduced density matrix element $(\rho_{ij})_{1,4} = 0 $, which is our case, $m$ and $n$ become irrelevant and $S(\rho_{ij}| \{B_k\} )$ depends only on $k$. Therefore, we only have to compare $S(\rho_{ij}| \{B_k\} )$ for $k=1/2$ and $k=0 $. \\
By the formulas in Ref. ~\cite{Ali}, for $k=1/2$ we have
\bae
& S(\rho_{ij}| \{B_k\} ) \equiv S_1 (\rho_{ij}) = \label{S1}   \\ \nonumber
& -\frac{1-\theta_1 }{2} \log_2 \frac{1-\theta_1 }{2} - \frac{1+\theta_1 }{2} \log_2 \frac{1+\theta_1}{2}
\eae
where
\be
\theta_1 =   \sqrt{ [(\rho_{ij})_{11} -(\rho_{ij})_{33} + (\rho_{ij})_{22} - (\rho_{ij})_{44} ]^2 + 4 |(\rho_{ij})_{23}|^2}
\ee
while for $k=0$ we have
\bae
& S(\rho_{ij}| \{B_k\} ) \equiv S_2 (\rho_{ij})  = \label{S2} \\ \nonumber
& -(1-p_0) \frac{1-\theta_2 }{2} \log_2 \frac{1-\theta_2 }{2} - (1-p_0) \frac{1+\theta_2 }{2} \log_2 \frac{1+\theta_2}{2} \\ \nonumber
& -p_0 \frac{1-\theta_3 }{2} \log_2 \frac{1-\theta_3 }{2} - p_0 \frac{1+\theta_3 }{2} \log_2 \frac{1+\theta_3}{2}
 \\ \nonumber
\eae
where $\ p_0 = (\rho_{ij})_{11}  + (\rho_{ij})_{33} $ and
\be \theta_2 =  \frac{| (\rho_{ij})_{22} - (\rho_{ij})_{44} |} {|(\rho_{ij})_{22} + (\rho_{ij})_{44}| } , \
\theta_3 =   \frac{| (\rho_{ij})_{11} - (\rho_{ij})_{33} |} {|(\rho_{ij})_{11} + (\rho_{ij})_{33}| }
\label{S3}
\ee
All we must do is take the minimum between (\ref{S1}) and (\ref{S2}):
\begin{equation}
\mbox{inf}_{\{B_k\}} S(\rho_{ij} | \{B_k\}) = \min\{ S_1, S_2 \}   \label{minimumqubit}
\end{equation}

\subsection{two-qutrit states} \label{qutrit}
As for the two-qutrit case, we have that the possible von Neumann measurements correspond to unitary rotations,
\be
B_0 = V \ket{0} \bra{0} V^\dag \ , B_1 = V \ket{1} \bra{1} V^\dag  \ , B_2 = V \ket{2} \bra{2} V^\dag
\ee
where now $V \in SU(3)$. \\
Unfortunately, to proceed forward in the computation of the discord, one cannot simply mimic the procedure described for qubits. The main difficulty is that no easy, explicit parametrization of $V \in SU(3)$ by 8 real parameters (the group dimension) can be found.~\cite{MacFarlane}.
We therefore must compute the discord \textit{numerically}. Our strategy is to minimize $ S(\rho_{ij} | \{B_k\} ) $ over a (large) set of randomly-generated unitary matrices~\cite{Zyczkowski}. More precisely, we generate a large ensamble of unitary matrices taken from the uniform distribution over the $SU(3)$ group manifold, evaluating $S(\rho_{ij}| \{B_k\} ) $ for each matrix. We then keep the minimum as our esteem of $\inf_{\{ B_k\} }S(\rho_{ij}| \{B_k\} )$. To be rigorous, this esteem is to be regarded as an upper bound: however, since we are generating a rather large set of random matrices we are confident that the bound is very stringent. \\
Alternatively, we can use the $SU(3)$ parametrization given in Ref.~\cite{Bronzan}. This allows to parameterize $SU(3)$ in terms of trigonometric functions of $8$ independent parameters, $3$ angles $\eta_1, \eta_2, \eta_3$ and $5$ phases $\alpha, \beta, \gamma, \rho, \sigma$. This parametrization makes it apparent that the phases $\rho$ and $\sigma$ are completely irrelevant for the computation of the discord, since orthonormal projectors (von Neumann measurements) are independent of the choice of such phases. This method has the advantage that it is based on a more transparent parametrization of von Neumann measurements. Again, we generate a large ensemble of unitary matrices find the minimum of $S(\rho_{ij}| \{B_k\} ) $ \\
In all cases under study, the two methods applied led to the same results, which provides us with full confidence on their reliability.

\end{document}